\documentclass[reprint,amsmath,amssymb,aps,pre,superscriptaddress,a4paper]{revtex4-2}

\usepackage{graphicx}
\usepackage[colorlinks=true,allcolors = blue]{hyperref}

\usepackage{newtxtext,newtxmath}
\usepackage{siunitx}
\newcommand{\Eq}[1]{Eq.~\eqref{#1}}

\newcommand{\Fig}[1]{Fig.~\ref{#1}}

\newcommand{\Figref}[1]{Fig.~\ref{#1}}

\newcommand{\latin}[1]{{\itshape #1}}
\newcommand{\etal}{\latin{et al.}\  }

\newcommand\rev[1]{\textcolor{black}{#1}}

\begin{document}

\title{Estimating the population-level effects of non-pharmaceutical interventions when transmission rates of COVID-19 vary by orders of magnitude from one contact to another
}

\author{Richard P. Sear}
\email{r.sear@surrey.ac.uk}
\affiliation{School of Mathematics and Physics, University of Surrey, Guildford, GU2 7XH, United Kingdom}
\homepage{https://richardsear.me/}

\begin{abstract}
Statistical physicists have long studied systems where the variable of interest spans many orders of magnitude, the classic example is the relaxation times of glassy materials, which are often found to follow power laws. A power-law dependence has been found for the probability of transmission of COVID-19, as a function of length of time a susceptible person is in contact with an infected person. This is in data from the United Kingdom’s COVID-19 app. The amount of virus in infected people spans many orders of magnitude. Inspired by this I assume that the power-law behaviour found in COVID-19 transmission, is due to the effective transmission rate varying over orders of magnitude from one contact to another. I then use a model from statistical physics to estimate that if a population all wear FFP2/N95 masks, this reduces the effective reproduction number for COVID-19 transmission by a factor of approximately nine.
\end{abstract}


\maketitle


\section{The transmission of diseases across the air and Non-Pharmaceutical Interventions}

\begin{figure}[b!]
  \centering
  \includegraphics[width=0.95\linewidth]{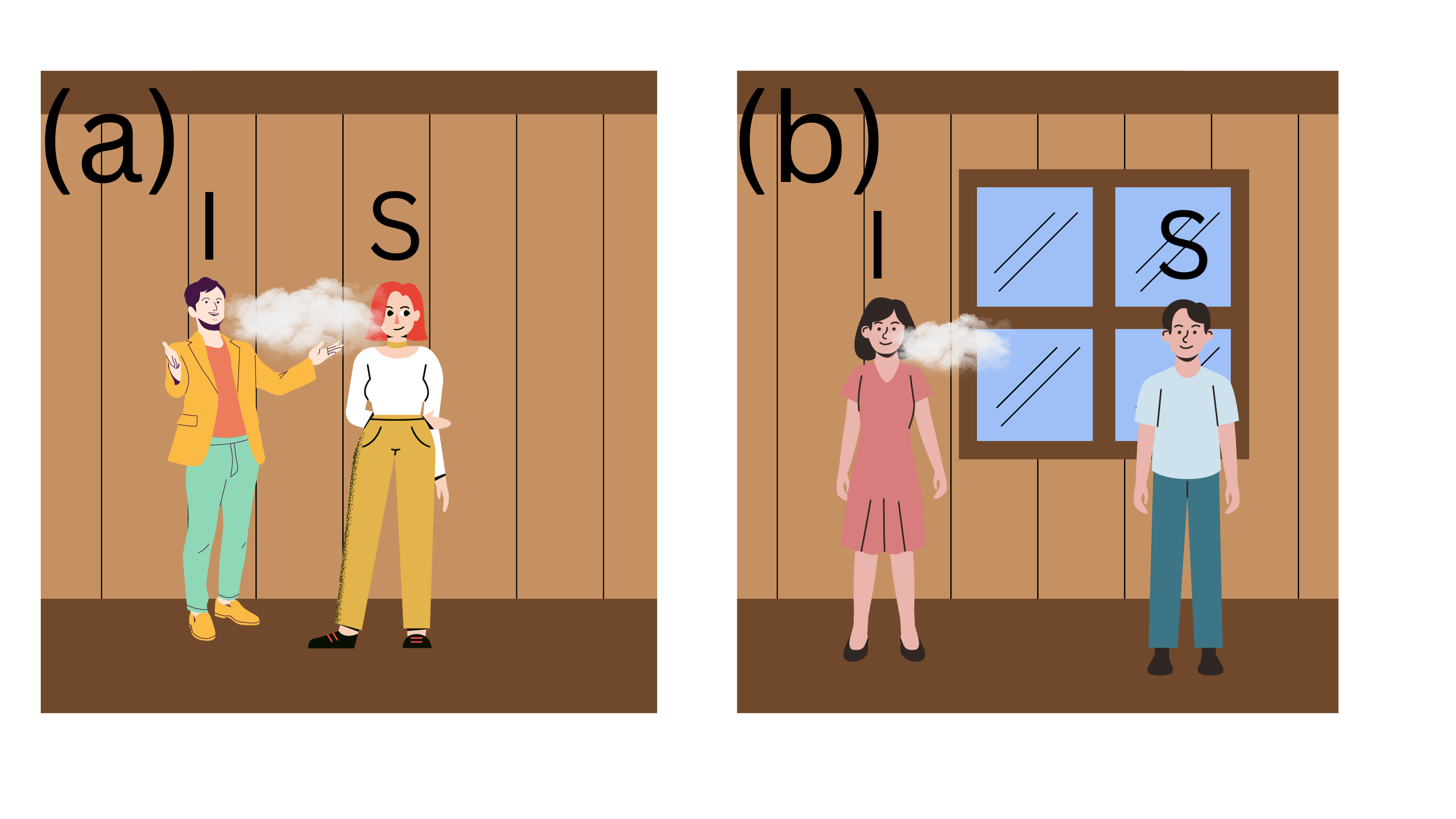}
  \caption{Pair of schematics showing two possible contacts between an infected person (I, left) and a susceptible person (S, right). The evidence is that the transmission of COVID-19 is at least predominantly across the air \cite{greenhalgh2021,pohlker2023,abkarian2020,adenaiye2021,tan2023,jia2022,archer2022}: an infected person breathes out the SARS-CoV-2 virus in an aerosol of droplets of mucus that contain the virus. The droplets are then carried by air currents towards a susceptible person, who inhales them and becomes infected \cite{pohlker2023,greenhalgh2021,abkarian2020,jia2022}. In panel (a) the rate of transmission $r$ is high because an infected person with a high viral load is close to and talking with a susceptible person with high genetic susceptibility to infection. In panel (b) the transmission rate $r$ is much lower. There the infected person has much less virus (lower viral load) and the susceptible person is across a well ventilated room from the infected person. }
  \label{fig:two_people}
\end{figure}

The standard model for the transmission of an airborne infectious disease such as COVID-19 postulates that transmission is a random process where a susceptible person takes up an infectious dose of the virus \cite{haas2015,szeto2010}, and becomes infected. This transmission takes place during a contact between an infected and a susceptible person. The contact is a time period of length $t$, during which the infected and susceptible person are close, for example in the same room. Figure \ref{fig:two_people} is a schematic illustrating the mechanism of airborne transmission of COVID-19.


If we assume standard Poisson statistics for the inhalation of the virus and becoming infected, we have that the probability of transmission increases with contact time $t$ as
\begin{equation}
    P_T(t)=1-\exp(-rt)\sim rt~~~~~ rt\ll 1
    \label{eq:Poisson}
\end{equation}
with $r$ the rate constant for transmission. For airborne diseases such as COVID-19, as well as flu and tuberculosis (TB), \Eq{eq:Poisson} is usually called the Wells-Riley model \cite{szeto2010,haas2015,jones2021}. 
In \Eq{eq:Poisson} we have expanded out the exponential to show that at low infection probabilities, the probability of transmission is predicted to increase linearly with time.


Non-pharmaceutical interventions (NPIs), such as mask wearing typically work via reducing the inhalation rate of virus \cite{greenhalgh2024}. For example, wearing an FFP2/N95 mask (respirator) reduces the inhaled dose by a factor of approximately 10 \cite{bagheri2021,robinson2022,zoller2021,duncan2021}.
We want to know by how much does this reduce the probability of transmission $P_T$? 

If we are recommending an NPI for a population then we will not know what the contact time $t$ or transmission rate $r$ is. Both may vary widely from one contact to another. This leaves us with the difficult problem of assessing the effectiveness of an intervention when we know neither the transmission rate nor the contact time. To make progress I combine two things. The first is a method to model systems with high variability, here in the transmission rate $r$, which I take from statistical physics \cite{johnston2006,newman2005,clauset2009,lindsey1980,wu2016}. The second is a large data set on COVID-19 transmission, taken from the United Kingdom's National Health Service (NHS) app \cite{kendall2023,ferretti2024}. 

\begin{figure}[tbh!]
  \centering
  \includegraphics[width=0.92\linewidth]{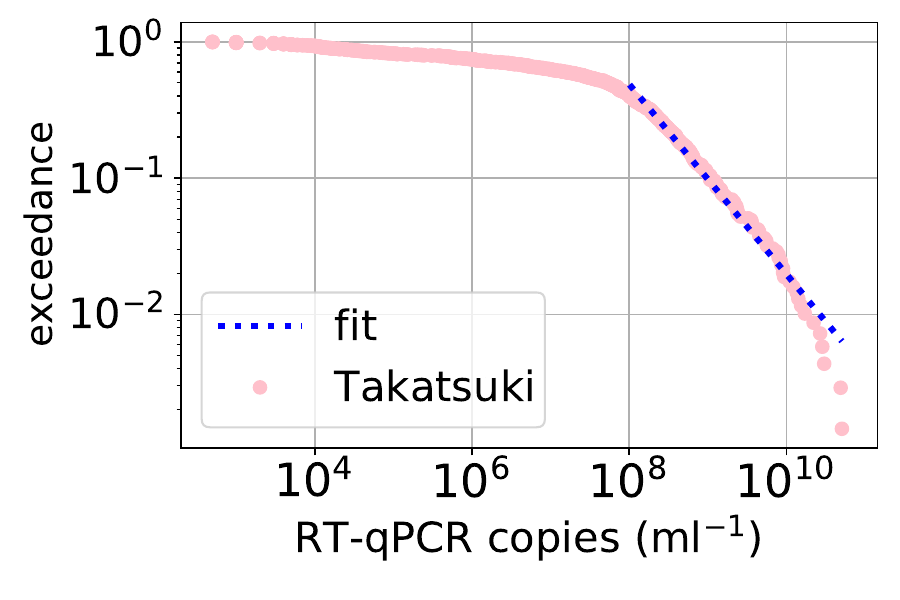}
  \caption{Plot of the exceedance function for the estimated concentration of RNA fragments of the COVID-19 genome, as measured by RT-qPCR. Points are from the study of 
  Takatsuki~\etal\cite{takatsuki2023}. The line is a power-law fit to the tail of the exceedance with an exponent 
  $-0.70$. }
  \label{fig:winnett}
\end{figure}

\section{Motivation for expected transmission rates to be highly variable}

I start by motivating my assumption that the transmission rate varies over orders of magnitude.
Transmission is complex \cite{peng2022}. There are many possible factors that could introduce variability into the rate $r$, both from human biology, and from the transport of the virus across the air from one person to another, see \Figref{fig:two_people}. We can break down factors that influence the transmission rate $r$ into three: 1) infectiousness of the infected person, for example how much virus is in their respiratory tract; 2) transport across the air; 3) susceptibility of person infected. 
We start with 1) infectiousness.

\subsection{Infectiousness}

In \Figref{fig:winnett} I have plotted the exceedance for the distribution of viral loads of people infected with COVID-19. The exceedance $E(v)$ is the probability that the viral load $v$ exceeds the value $v$. The data is from
Takatsuki~\etal\cite{takatsuki2023}. Note that the viral load is measured by a procedure that quantifies the amount of a fragment of COVID-19's genome, via reverse transcription quantitative polymerase chain reaction (RT-qPCR). This is not the same as a measure of the amount of infectious virus present \cite{puhach2023}.
However Killingley \etal\cite{killingley2022} found an approximately linear relationship between RT-qPCR measurements and a measure of how much viable virus there is, see 
Appendix \ref{app:linear}. 


The dynamic range of viral loads (as measured by RT-qPCR) is huge: 
8 orders of magnitude for the Takatsuki \etal\cite{takatsuki2023} data. Presumably, those with the highest viral load are most infectious \cite{puhach2023}, but we lack direct evidence of this.

The large viral-load tails of the exceedance is reasonably well fit by a power law with an exponent of $-0.7$. This corresponds to probability density functions for viral loads that scale as $r^{-1.7}$, for large viral loads. It is an empirical observation that data that spans many decades is often well approximated by a power law, in particular in the tail of large values, as here \cite{newman2005,clauset2009}. This applies to data in many complex systems, from the economy to earthquakes \cite{newman2005,clauset2009}. In complex systems such as viral infection or the economy, the power law is a simple but approximate empirical model, which can be used to make simple predictions. This is how we will use power law models here.


\subsection{Fluid flow}

Now we turn to how the virus is transported across the air, factor 2. This transport of infectious particles is most efficient when the infected and susceptible persons are close to each and facing each other \cite{abkarian2020,jones2021,jia2022,peng2022}. Abkarian and coworkers \cite{abkarian2020} found that for two people close ($\lesssim 0.5$ to $\SI{1}{\metre}$) and facing each other, that one will directly inhale some of the other's breath. This would be the case for example because they are talking to each other. There, perhaps up to $\sim 10\%$ of the air inhaled by one person, could be air recently exhaled by the other person. This fraction roughly decays as one over the separation up to distances around $\SI{2}{\metre}$, beyond which directly exhaled air will mix with room air \cite{abkarian2020,giri2022,bourouiba2021,bourouiba2021b,morawska2024,jia2022,peng2022}.

For a large well ventilated room, the fraction of room air that has been breathed out by one person (here the infected person) will be much less than 1\%. So when a susceptible person shares a room with an infected person, the air breathed out by the infected person could constitute anywhere between much less than 1\%, and approximately 10\% of the air the susceptible person is breathing in.

The air that we breathe out is saturated with water vapour, at 100\% relative humidity (RH).
For transmission across a room, the air of an infected person's breath mixes with room air as it crosses the room, and before being breathed in by a susceptible person. As the breath air mixes with room air, the RH drops sharply. From the near 100\% RH of breath to the typically close to 50\% RH of a room. This partial drying out inactivates a variable fraction of the virus SARS-CoV-2 \cite{oswin2022,haddrell2023}.
So the combination of the dilution of the virus-carrying breath of an infected person, and viral inactivation, can cause the concentration of active virus in the air to vary over orders of magnitude.

\subsection{Susceptibility}

There is evidence for variability in susceptibility to infection, \rev{susceptibility may greatly from one person to another. Part of this is due to} genetic differences between one person and another \cite{augusto2023,hill2012,vandermade2022}. \rev{In addition, susceptibility may vary with age \cite{viner2021}, and it will be higher in those with immunodeficiencies \cite{shoham2023,paris2023}. Vaccination reduces but does not eliminate susceptibility, by an amount that depends on vaccine and on virus variant \cite{kaplonek2023}, so varying vaccination status also introduces variability into the susceptibility to infection.}

\begin{figure}[tbh!]
  \centering
  \includegraphics[width=1.05\linewidth]{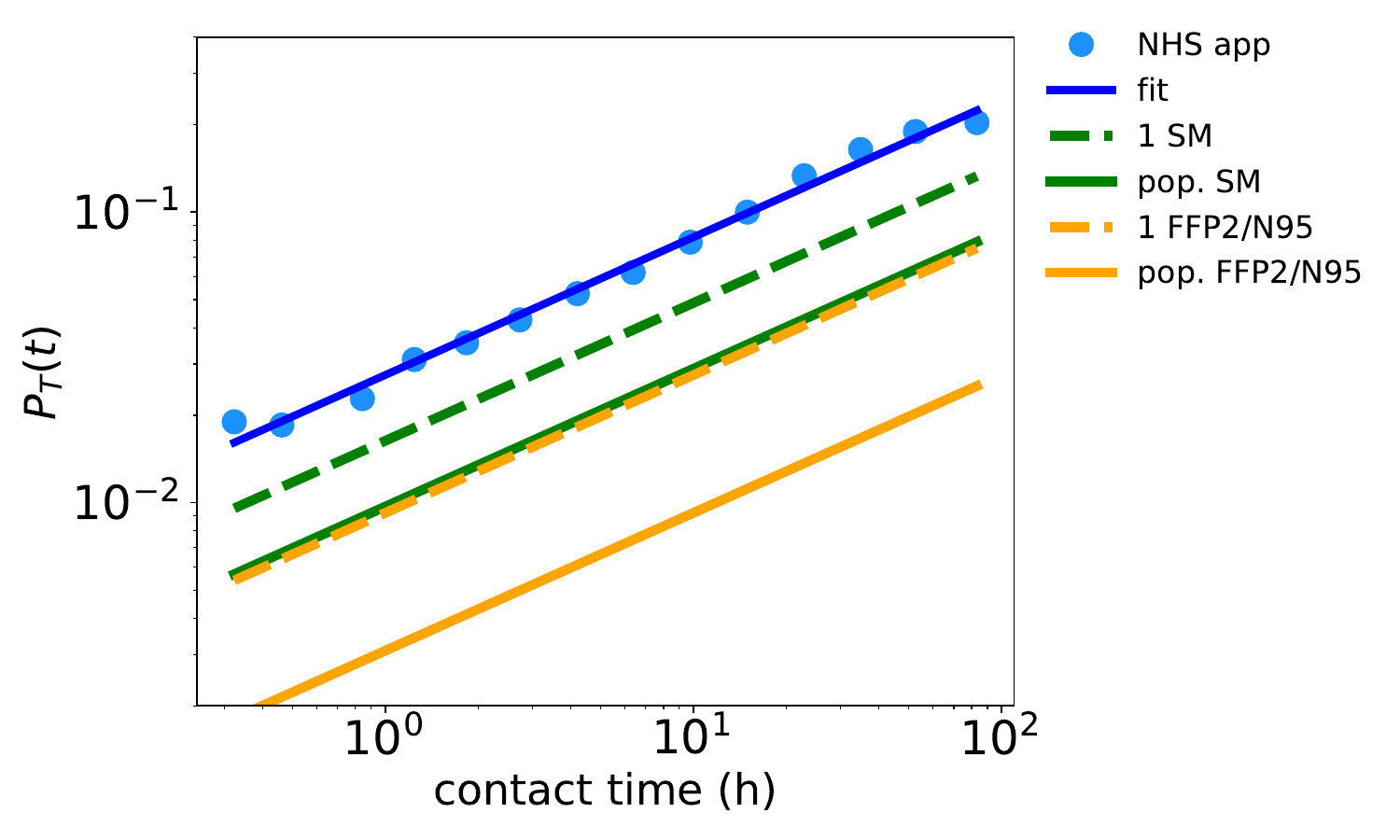}
  \caption{Plot of the probability of transmission of COVID-19, as a function of the length of contact time. The blue points are data from the work of Ferretti \etal\cite{ferretti2024}, using data from UK's COVID-19 app \cite{kendall2023}. Fit (blue) is obtained by fitting a straight line to logs of the data, and fit parameters are slope (exponent) $0.47\pm0.01$ and intercept $-3.6$.
  $R^2=0.996$. Ferretti~\etal\cite{ferretti2024} also fitted  a power law to data and found the same value for the exponent $\beta$.
  Data has dynamic range of three decades in time (minutes to tens of hours). Also shown are predictions for $P_T$ when: one susceptible person wears a surgical mask (green dashed), one person wears an FFP2/N95 (orange dashed), when the whole population wears surgical masks (solid green), and when the population wears FFP2/N95s (solid orange).}
  \label{fig:ferretti}
\end{figure}

\subsection{Overdispersion}

Finally, we comment that the \rev{that the transmission of COVID-19  is known to be overdispersed\cite{chen2021,wegehaupt2023}.} 
Overdispersion \rev{is a term in epidemiology that} refers to the width of the distribution of $N_{I}$: the number of people infected by one infectious person. 
The standard model in epidemiology is that $N_I$ follows a Poisson distribution. But if the distribution of $N_I$ is broader than a Poisson then the disease transmission is said to be overdispersed \cite{wegehaupt2023}

A broad distribution of infectiousness in infected people, for example due to a broad distribution of viral loads, provides a natural explanation for overdispersion. Or to reverse the argument, the observed overdispersion is consistent with highly variable effective transmission rates. \rev{However, there are other factors in transmission, for example the number of contacts of a person is itself overdispersed in the sense of being broader than a Poisson distribution \cite{mossong2008}. So any variability in transmission rate for a single contact is likely to be only one factor behind overdispersion.}

\section{Data on the probability of infection as a function of time}

So we expect that the transmission rate $r$ varies over orders of magnitude. We now turn to data from the United Kingdom's National Health Service (NHS) COVID-19 app \cite{kendall2023}.
During the COVID-19 pandemic, this app ran on users' mobile phones and used Bluetooth to detect other users' mobile phones when they were nearby. So it could estimate the length of a contact between two app users. Users also reported infections via the app. So both the duration of contacts and the probability of transmission can be estimated. See Kendall \etal\cite{kendall2023} and Ferretti~\etal\cite{ferretti2024} for further details.  

The data from the NHS app was analysed by Ferretti~\etal\cite{ferretti2024}, resulting in a plot of the probability of infection as a function of time, see \Figref{fig:ferretti} and Ferretti~\etal\cite{ferretti2024}. The measured probability of transmission $P_T$ is plotted in \Figref{fig:ferretti}. It is a particular population average: an average over the population of NHS app users in the UK during the period of operation of the app during the pandemic. This is data from 7 million contacts and 240,000 positive COVID-19 tests \cite{ferretti2024}.

As Ferretti and coworkers \cite{ferretti2024} found, the data is well fit by a power law with exponent that is close to one half. By well fit we mean that the fractional error (residual) in the fit is typically 10\% or less. 

\section{Wells-Riley model for infection probabilities in the presence of highly variable transmission rates}

We now return to the model for transmission, \Eq{eq:Poisson}. This gives the probability of transmission when there is one transmission rate $r$.
If there is a distribution $p(r)$ of the rates $r$, the probability of transmission is the
Laplace transform of the distribution $p(r)$
\begin{equation}
  P_T(t)=
    1-\int_0^{\infty}  p\left(r\right)
    \exp(-rt){\rm d}r
    \label{eq:pi_laplace}
\end{equation}
Here the probability density function $p(r)$ gives the probability that in a population a contact between an infected and a susceptible person, has a transmission rate $r$. 


If the distribution $p(r)$ is sharply peaked at some characteristic rate $r^*$, then $P_T(t)$ is close to the simple Poisson statistics model of \Eq{eq:Poisson}. This is linear at short times  --- not what is observed in \Figref{fig:ferretti}. In fact any relatively narrow distribution (no heavy/fat tails) of rates, such as the beta distribution used by Haas and coworkers and by others \cite{haas2015,szeto2010}, will give a $P_T(t)$ that is linear at short times.

\subsection{Broad (fat-tailed) distributions of transmission rates}

Very broad distributions are often modelled as power laws \cite{newman2005,clauset2009,stumpf2012}. This motivates us to consider a power law distribution of transmission rates $r$ with an exponent $x$,
\begin{equation}
p(r) \sim r^{-x}~~~~~r\gg r_{MED}
\label{eq:pofr}
\end{equation}
for $r_{MED}$ the median rate. Here we are only assuming the rates follow a power law in the large rate tail of the distribution (as seen for the viral loads in \Figref{fig:winnett}).

Johnston \cite{johnston2006} has studied the Laplace transforms of distributions $p(r)$ with power laws. 
Applying his results here, we find that distributions $p(r)$ with power law tails with $x<2$ lead to a power law probability of transmission (at the low probabilities of transmission of \Figref{fig:ferretti})
\begin{equation}
P_T(t)\sim (r^*t)^{\beta}~~~\mbox{at short times}
\label{eq:PIstretch}
\end{equation}
with $r^*$ a rate scale parameter and $\beta=x-1$. 

The data of Ferretti \etal\cite{ferretti2024} is consistent with a power law with exponent $\beta=0.47$. Therefore their data for COVID-19 transmission is consistent with the Wells-Riley model for transmission, with a power law distribution of transmission rates with exponent $x=\beta+1=1.47$. 

Johnston \cite{johnston2006} studied complete $P_T$'s that were stretched exponentials, i.e.,
\begin{equation}
P_T(t)=1-\exp\left[-(r^*t)^{\beta}\right]
\label{eq:PIstretch1}
\end{equation}
which give the power law behaviour in \eqref{eq:PIstretch} at short times/low $P_T$.
See 
Appendix \ref{app:laplace} 
for the full distribution $p(r)$ for $x=3/2$ that gives rise to a stretched exponential. Note that as we only have data at low transmission probabilities, we can infer the large rate tail of $p(r)$ but have little information on the behaviour of $p(r)$ at low rates.



It is worth noting that the Wells-Riley model assumes that the probability of remaining uninfected decays exponentially, equivalent to assuming that the transmission rate $r$ is not changing with time. This is clearly an approximation. The rate should vary whenever either the infected or susceptible person moves, for example farther apart or closer together. And the measured viral load varies from one day to the next \cite{killingley2022,jones2021,kissler2021}.
However, the exponential dependence for a single contact should be a reasonable approximation so long as for most contacts there is fairly well defined timescale for transmission.

A final point to note is that the rate of increase of $P_T(t)$ with time is decreasing, its first derivative with respect to time scales as $t^{\beta-1}$. As we have known since the work of Proschan in the 1960s \cite{proschan1963} (on different systems), this is characteristic of systems where each one (here each contact) obeys Poisson statistics, but each system has a different rate. As Proschan \cite{proschan1963} pointed out, the reason for this is simple. At early times the rate of increase of $P_I$ is dominated by contacts with high transmission rates, so at these early times, $P_T$ increases rapidly. But at longer times, transmission has already occurred for these high-transmission-rate contacts, so at these longer times they do not contribute to the increasing $P_T$. At these longer times, transmission is dominated by contacts with slower rates, and so the rate of increase of $P_T$ with time is slower.

\section{Model for the effect of NPIs}

Now that we have a model for the distribution of transmission rates, \Eq{eq:pofr}, we can develop a simple predictive model for the effect of an NPI. We assume that an NPI removes/filters out a fraction $(1-f)$ of the virus and so scales {\em all} transmission rates $r$ by a factor $f$, i.e.,
\begin{equation}
r \xrightarrow{\text{NPI}}fr
\label{eq:pofr_scaled}
\end{equation}
This then simply scales the rate-scale parameter $r^*$ by a factor $f$ and, then using \Eq{eq:PIstretch}
\begin{equation}
    P_T^{NPI}(t)\sim (fr^*t)^{\beta} \sim f^{\beta}P_I(t)
    ~~~~\mbox{at short times, i.e., small $P_T$}
    \label{eq:PI2}
\end{equation} 
We predict that when an NPI that reduces the transmission rates by a factor $f$ is applied, the probability of infection scales by a factor $f^{\beta}$.
For the population of NHS app users, $\beta\simeq 1/2$.
As $\beta<1$ this is a sublinear reduction in risk, you need to reduce the expected dose by a factor of approximately four to reduce the risk by a factor of two. 

It is worth noting that to make predictions for the effect of an NPI, we need a model for transmission, which necessarily makes assumptions. An empirical fit such as the power-law fit in \Fig{fig:ferretti} fixes a parameter value ($\beta$) in a fit but the fit itself is not enough. We need to assume: 1) the transmission probability for an individual contact obeys Poisson statistics, and 2) we are observing an average of a population of (independent) individual contacts where the rate $r$ varies from one contact to another. Then the fit allows us to estimate the distribution of these rates. We then make assumption 3) that the NPI reduces the expected transmission rate of all contacts by a factor of $f$. Only then can we make an approximate prediction for the effect of an NPI.


\subsection{Estimate for reduction in risk due to FFP2s/N95s and surgical masks}

One NPI is the wearing of masks such as N95 or FFP2 masks \cite{greenhalgh2024}.
N95 masks have been assigned a protection factor of 10, i.e., $f=0.1$ \cite{duncan2021}. The FFP2 standard is for a filtration efficiency of 92\% as worn \cite{zoller2021,robinson2022}. Taking an FFP2/N95 to reduce the dose by a factor of $f=0.1$  we can estimate $P_T(t)$ for a person who is a member of a population equivalent to that of the NHS app users, when this person wears an FFP2/N95. This is the orange dashed line in \Figref{fig:ferretti}. The prediction is that for contacts of all durations, the reduction in risk due to wearing an FFP2/N95 is by a factor of approximately three.

If the entire population dons FFP2/N95 masks the air is filtered twice, on exhaling and inhaling. The filtering efficiency should be approximately the same in both directions, so now $f=0.1^2=0.01$. Then the reduction in risk is by approximately a factor of $1/0.01^{0.47}\simeq 9$. This is the solid orange line in \Figref{fig:ferretti}.

Surgical masks are not only far less effective than FFP2/N95s, they are highly variable \cite{robinson2022,oberg2008,duncan2021}. The FFP2 and N95 standards include requirements for filtration efficiency as worn but standards for surgical masks do not. Measured reductions in inhalation rate are by factors of $2.5$ to $6.9$ in the work of Oberg and Brosseau \cite{oberg2008}, and $1.7$ to $3.6$ in the work of Duncan~\etal\cite{duncan2021}. If we take a factor of three then this gives a reduction in risk of $(1/3)^{0.47}\simeq 0.6$. So we predict one person wearing a surgical mask gives a reduction in infection probability of about 40\%, while a whole population wearing surgical masks reduces transmission by about 60\%. These predictions are shown as the dashed and solid green lines in \Figref{fig:ferretti}.

\subsubsection{Effect of NPI on effective reproduction number}

In epidemiology, transmission of an infectious disease is often quantified by the effective reproduction number, which is the average number of infections caused by a single infected person \cite{vegvari2022,vanderdriessche2017,cori2013}. It is usually denoted by $R$ (or $R_t$). For an NPI that reduces the inhaled dose by a constant fraction $f$ then as the transmission rate is predicted to drop by the constant fraction $f^{\beta}$, then the effective reproduction number $R$ is also predicted to scale by the same factor, i.e.,
\begin{equation}
R \xrightarrow{\text{NPI}}f^{\beta}R
\label{eq:R_scaled}
\end{equation}
So if a population dons FFP2/N95 masks, then as above $f=0.01$ and so $R$ should be reduced by a factor of about nine.





\subsection{Testing the prediction}


There have been both observational studies and randomised controlled trials (RCTs) of the number of COVID-19 infections in those wearing masks, see the recent review of Boulos~\etal~\cite{boulos2023}.
However, there is considerable scatter in the sizes of the effects found in these studies \cite{boulos2023}. The studies all suffer from some combination of methodological problems \cite{boulos2023} and small sample size (so large statistical error bars) \cite{boulos2023}. In addition, the analysis of the data is subtle and controversial \cite{baryam2023}. 

So, the RCTs and observational studies neither disprove the prediction nor strongly support the prediction made here. RCTs of mask use are challenging. They require a large population (to obtain small statistical error bars) to be split into two, with one half always wearing FFP2/N95 masks during any exposure, while otherwise behaving identically to the unmasked control group. As they are so challenging, we may have to rely on predictions such as that obtained here.

\section{Conclusion}

In conclusion, I have predicted that for a population like that of the UK, wearing FFP2/N95-type masks should reduce the effective reproduction number $R$ by a factor of nine. This relies on the COVID-19 app data, where there is an approximately square-root dependence of the probability of transmission \cite{ferretti2024}, on the length of time of a contact between an infected and a susceptible person. This sublinear dependence on time can be interpreted as due to the transmission rate varying widely between one contact and another. Roughly speaking, the probability of transmission after, for example, a contact of 2 hours, is set by the fraction of contacts where the transmission rate is high enough to drive transmission in 2 hours or less. Then mask use reduces the inhaled doses and so scales up the time needed for transmission, such that a smaller fraction of contacts are long enough for transmission.

I have focused on one NPI for airborne transmission: masks/respirators. Another NPI is improving indoor air quality, as Morawska~\etal \cite{morawska2024} have recently called for.
This work suggests that, for example, doubling the rate of turnover of air should reduce transmission by approximately $1-1/\sqrt{2}\sim 30\%$. This estimate is highly approximate but as obtaining population-level data on the effect of NPIs is so hard to do, this estimate may still be useful.

\noindent
{\bf Acknowledgement}
This work relies on the data collected by the NHS COVID app \cite{kendall2023,ferretti2024}, so thank you to those who analysed the data and all who used this app. I also acknowledge that this work also relied on the data of 
Takatsuki \etal \cite{takatsuki2023} and Killingley \etal \cite{killingley2022}.

\noindent
{\bf Supplemental Material}
This is a Python Jupyter notebook that performs all the calculations in this work. This is available as XXX.

\appendix

\section{Approximately linear relationship between a measurement of viable virus and qPCR measurement}
\label{app:linear}

\begin{figure}[tbh!]
  \centering
  \includegraphics[width=0.95\linewidth]{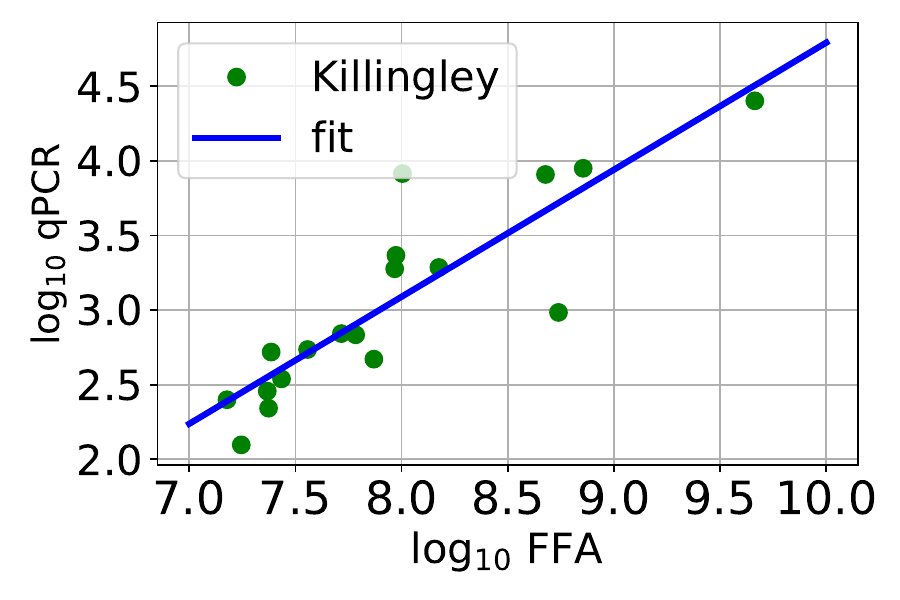}
  \caption{Plot of $\log_{10}($qPCR$)$ as a function of $\log_{10}($FFA$)$, data is from Killingley~\etal\cite{killingley2022}. Their extended data Fig.~2(b), from throat swabs of people infected with COVID-19.}
  \label{fig:killingley}
\end{figure}

Killingley \etal\cite{killingley2022} find an approximately linear relationship between the quantity of viable virus and a qPCR measurement, for samples taken from the throat. Viable virus is measured by a focus-forming assay (FFA) \cite{killingley2022}. Their data and a fit are in \Figref{fig:killingley}. The fit of a straight line to log data gives an exponent of $0.89\pm 0.13$, so close to a simple linear dependence. But note that the relationship may vary through the time course of an infection \cite{ke2021,kissler2021,jones2021}.

There is some evidence for a correlation between transmission, and  the viral load some time after transmission, see the review of Puhach \etal~\cite{puhach2023}. Note that we almost never know the viral load at the time transmission occurs, because the fact that transmission has occurred is only determined later. The viral load varies greatly during the course of an infection \cite{puhach2023}.


\begin{figure}[tbh]
  \centering
  \includegraphics[width=0.95\linewidth]{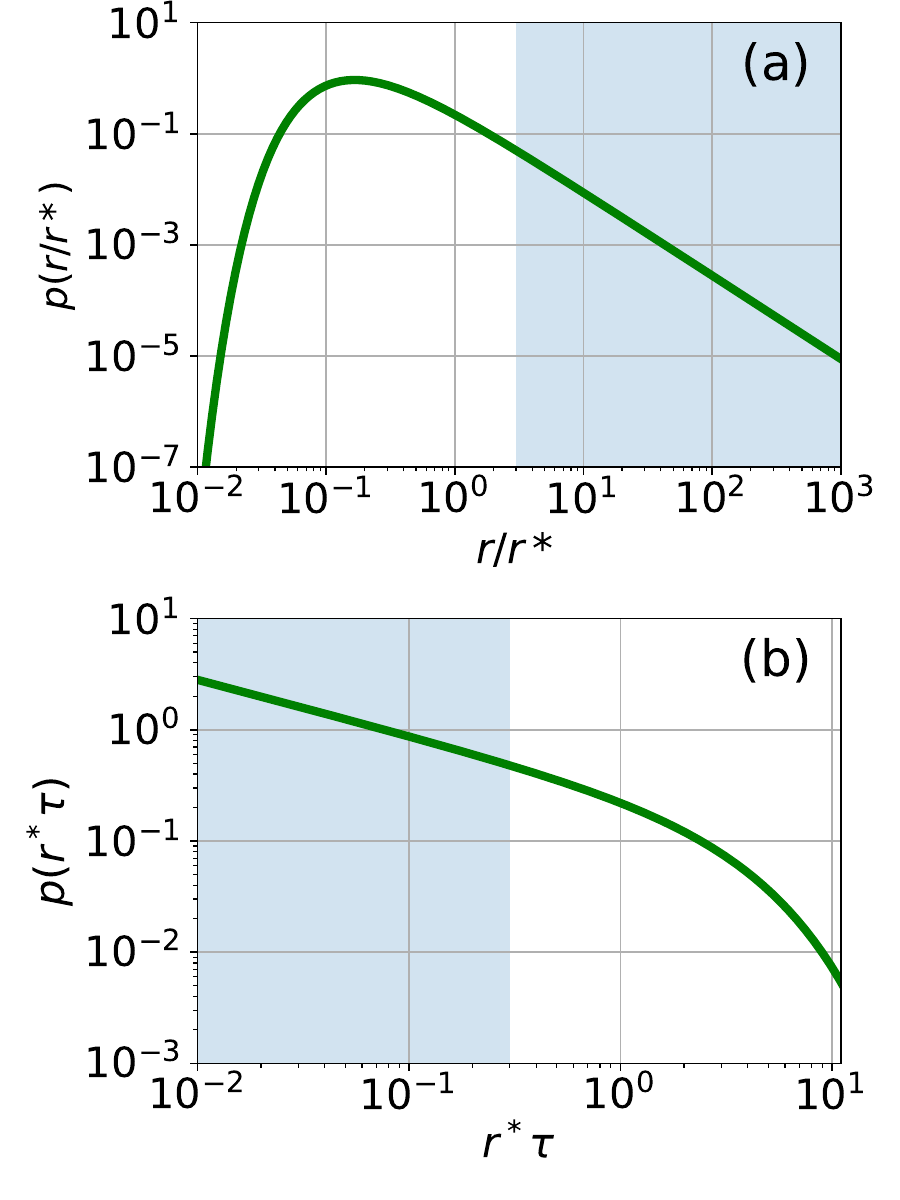}
  \caption{(a) Plot of the probability distribution function of transmission rates, $p(r/r^*)$, \Eq{eq:laplace_sqrt}, that yields a stretched exponential $P_T(t)$ with exponent $\beta=1/2$ \cite{johnston2006}. (b) Plot of the probability distribution function of transmission timescales $\tau$, $p(r^*\tau)$, \Eq{eq:laplace_sqrt_tau}, that yields a stretched exponential $P_T(t)$ with exponent $\beta=1/2$ \cite{lindsey1980,wu2016}. In both (a) and (b) the shaded regions are the only parts of the distribution relevant to the range of times (tens of minutes to tens of hours) for which the NHS app provides data.}
  \label{fig:laplace}
\end{figure}

\section{A stretched-exponential function with an exponent of one half is the Laplace transform of a distribution of rates with a large-rate tail scaling as $r^{-3/2}$}
\label{app:laplace}

The probability density function $p(s)$ that produces a stretched exponential $\exp\left[(-r^*t)^{\beta}\right]$ with exponent $\beta=1/2$ is \cite{johnston2006}
\begin{equation}
    p(s)=\frac{\exp\left(-1/s\right)}{\sqrt{4\pi s^{3}}}
    \label{eq:laplace_sqrt}
\end{equation}
with $s=r/r^*$, and $r^*$ a characteristic rate. This $p(s)$ is plotted as \Figref{fig:laplace}(a). The median reduced rate $s$ is $s_{MED}\simeq 1.1$ so the median rate $r_{MED}\simeq 1.1r^*$ \cite{johnston2006}. 

Note the power law behaviour $s^{-3/2}$, for \Eq{eq:laplace_sqrt} when $s\gg 1$. So as Ferretti~\etal\cite{ferretti2024}'s data is consistent with the low-infection-probability/short-time part of a stretched exponential with exponent near one half, it is consistent with a distribution of rates with a large-rate tail scaling approximately as $r^{-3/2}$. Ke~\etal\cite{ke2021} showed that a stretched exponential provides a good fit to data on infecting cell cultures as  a function of (PCR-measured) viral load, with exponent $\beta$ values similar to here, i.e., around one half.

However, note that the $p(r)$ in \Eq{eq:laplace_sqrt} has a small-rate cutoff at $s$ of order one, i.e., there are few rates much below the median rate, as we can see in \Figref{fig:laplace}(a). The lack of rates much below the median is why the stretched exponential tends to one for times longer than $1/r^*$. In contrast, the infection probability of Ferretti~\etal is only $P_T\lesssim 1/4$ at around 100 hours.

So although the large rate tail of the distribution of \Eq{eq:laplace_sqrt} and \Figref{fig:laplace}(a) (large rate tail is shaded in the figure) is consistent with the experimental data, presumably the low rate parts differ. The fact that Ferretti~\etal have maximum values of $P_T$ around one quarter not one within the tens of hours they measure implies that the effective transmission rate for three quarters of contacts is somewhere in the range zero to $10^{-2}$h$^{-1}$. Finally, note that what is measured here is {\em reported} (via the app) infections, so for example if a person as infected but showed few or no symptoms they may not realise they are infected and so not report it. See Ferretti~\etal\cite{ferretti2024} for a discussion.


This is all for a distribution of rates. 
We can also write $P_T$ as a distribution not of transmission rates $r$ but of transmission timescales $\tau$, as
\begin{equation}
  P_T(t)=\int_0^{\infty}  p\left(\tau\right)
    \left[1-\exp(-t/\tau)\right]{\rm d}\tau
    \label{eq:pi_laplace_tau}
\end{equation}
Then an exponent $\beta=1/2$ corresponds to the distribution of transmission timescales \cite{lindsey1980,wu2016}
\begin{equation}
    p(q)=\frac{\exp\left(-(1/4)q\right)}{\sqrt{4\pi q}}
    \label{eq:laplace_sqrt_tau}
\end{equation}
for $q=r^*\tau$, i.e., the transmission timescale for a contact $\tau$ in units of the characteristic timescale $1/r^*$. 

We have plotted $p(q)$ in \Figref{fig:laplace}(B). Note the broad distribution of characteristic timescales $\tau$. As we should expect, whether we view transmission is being due to a distribution of transmission rates or timescales, then the distributions we find are very broad.


\begin{thebibliography}{52}%
\makeatletter
\providecommand \@ifxundefined [1]{%
 \@ifx{#1\undefined}
}%
\providecommand \@ifnum [1]{%
 \ifnum #1\expandafter \@firstoftwo
 \else \expandafter \@secondoftwo
 \fi
}%
\providecommand \@ifx [1]{%
 \ifx #1\expandafter \@firstoftwo
 \else \expandafter \@secondoftwo
 \fi
}%
\providecommand \natexlab [1]{#1}%
\providecommand \enquote  [1]{``#1''}%
\providecommand \bibnamefont  [1]{#1}%
\providecommand \bibfnamefont [1]{#1}%
\providecommand \citenamefont [1]{#1}%
\providecommand \href@noop [0]{\@secondoftwo}%
\providecommand \href [0]{\begingroup \@sanitize@url \@href}%
\providecommand \@href[1]{\@@startlink{#1}\@@href}%
\providecommand \@@href[1]{\endgroup#1\@@endlink}%
\providecommand \@sanitize@url [0]{\catcode `\\12\catcode `\$12\catcode
  `\&12\catcode `\#12\catcode `\^12\catcode `\_12\catcode `\%12\relax}%
\providecommand \@@startlink[1]{}%
\providecommand \@@endlink[0]{}%
\providecommand \url  [0]{\begingroup\@sanitize@url \@url }%
\providecommand \@url [1]{\endgroup\@href {#1}{\urlprefix }}%
\providecommand \urlprefix  [0]{URL }%
\providecommand \Eprint [0]{\href }%
\providecommand \doibase [0]{https://doi.org/}%
\providecommand \selectlanguage [0]{\@gobble}%
\providecommand \bibinfo  [0]{\@secondoftwo}%
\providecommand \bibfield  [0]{\@secondoftwo}%
\providecommand \translation [1]{[#1]}%
\providecommand \BibitemOpen [0]{}%
\providecommand \bibitemStop [0]{}%
\providecommand \bibitemNoStop [0]{.\EOS\space}%
\providecommand \EOS [0]{\spacefactor3000\relax}%
\providecommand \BibitemShut  [1]{\csname bibitem#1\endcsname}%
\let\auto@bib@innerbib\@empty
\bibitem [{\citenamefont {Haas}(2015)}]{haas2015}%
  \BibitemOpen
  \bibfield  {author} {\bibinfo {author} {\bibfnamefont {C.~N.}\ \bibnamefont
  {Haas}},\ }\bibfield  {title} {\bibinfo {title} {Microbial dose response
  modeling: Past, present, and future},\ }\href
  {https://doi.org/10.1021/es504422q} {\bibfield  {journal} {\bibinfo
  {journal} {Environmental Sci. Tech.}\ }\textbf {\bibinfo {volume} {49}},\
  \bibinfo {pages} {1245} (\bibinfo {year} {2015})}\BibitemShut {NoStop}%
\bibitem [{\citenamefont {Sze~To}\ and\ \citenamefont
  {Chao}(2010)}]{szeto2010}%
  \BibitemOpen
  \bibfield  {author} {\bibinfo {author} {\bibfnamefont {G.~N.}\ \bibnamefont
  {Sze~To}}\ and\ \bibinfo {author} {\bibfnamefont {C.~Y.~H.}\ \bibnamefont
  {Chao}},\ }\bibfield  {title} {\bibinfo {title} {{Review and comparison
  between the Wells–Riley and dose-response approaches to risk assessment of
  infectious respiratory diseases}},\ }\href
  {https://doi.org/https://doi.org/10.1111/j.1600-0668.2009.00621.x} {\bibfield
   {journal} {\bibinfo  {journal} {Indoor Air}\ }\textbf {\bibinfo {volume}
  {20}},\ \bibinfo {pages} {2} (\bibinfo {year} {2010})}\BibitemShut {NoStop}%
\bibitem [{\citenamefont {Greenhalgh}\ \emph {et~al.}(2021)\citenamefont
  {Greenhalgh}, \citenamefont {Jimenez}, \citenamefont {Prather}, \citenamefont
  {Tufekci}, \citenamefont {Fisman},\ and\ \citenamefont
  {Schooley}}]{greenhalgh2021}%
  \BibitemOpen
  \bibfield  {author} {\bibinfo {author} {\bibfnamefont {T.}~\bibnamefont
  {Greenhalgh}}, \bibinfo {author} {\bibfnamefont {J.~L.}\ \bibnamefont
  {Jimenez}}, \bibinfo {author} {\bibfnamefont {K.~A.}\ \bibnamefont
  {Prather}}, \bibinfo {author} {\bibfnamefont {Z.}~\bibnamefont {Tufekci}},
  \bibinfo {author} {\bibfnamefont {D.}~\bibnamefont {Fisman}},\ and\ \bibinfo
  {author} {\bibfnamefont {R.}~\bibnamefont {Schooley}},\ }\bibfield  {title}
  {\bibinfo {title} {{Ten scientific reasons in support of airborne
  transmission of SARS-CoV-2}},\ }\href
  {https://doi.org/https://doi.org/10.1016/S0140-6736(21)00869-2} {\bibfield
  {journal} {\bibinfo  {journal} {Lancet}\ }\textbf {\bibinfo {volume} {397}},\
  \bibinfo {pages} {1603} (\bibinfo {year} {2021})}\BibitemShut {NoStop}%
\bibitem [{\citenamefont {P\"ohlker}\ \emph {et~al.}(2023)\citenamefont
  {P\"ohlker}, \citenamefont {P\"ohlker}, \citenamefont {Kr\"uger},
  \citenamefont {F\"orster}, \citenamefont {Berkemeier}, \citenamefont
  {Elbert}, \citenamefont {Fr\"ohlich-Nowoisky}, \citenamefont {P\"oschl},
  \citenamefont {Bagheri}, \citenamefont {Bodenschatz}, \citenamefont
  {Huffman}, \citenamefont {Scheithauer},\ and\ \citenamefont
  {Mikhailov}}]{pohlker2023}%
  \BibitemOpen
  \bibfield  {author} {\bibinfo {author} {\bibfnamefont {M.~L.}\ \bibnamefont
  {P\"ohlker}}, \bibinfo {author} {\bibfnamefont {C.}~\bibnamefont
  {P\"ohlker}}, \bibinfo {author} {\bibfnamefont {O.~O.}\ \bibnamefont
  {Kr\"uger}}, \bibinfo {author} {\bibfnamefont {J.-D.}\ \bibnamefont
  {F\"orster}}, \bibinfo {author} {\bibfnamefont {T.}~\bibnamefont
  {Berkemeier}}, \bibinfo {author} {\bibfnamefont {W.}~\bibnamefont {Elbert}},
  \bibinfo {author} {\bibfnamefont {J.}~\bibnamefont {Fr\"ohlich-Nowoisky}},
  \bibinfo {author} {\bibfnamefont {U.}~\bibnamefont {P\"oschl}}, \bibinfo
  {author} {\bibfnamefont {G.}~\bibnamefont {Bagheri}}, \bibinfo {author}
  {\bibfnamefont {E.}~\bibnamefont {Bodenschatz}}, \bibinfo {author}
  {\bibfnamefont {J.~A.}\ \bibnamefont {Huffman}}, \bibinfo {author}
  {\bibfnamefont {S.}~\bibnamefont {Scheithauer}},\ and\ \bibinfo {author}
  {\bibfnamefont {E.}~\bibnamefont {Mikhailov}},\ }\bibfield  {title} {\bibinfo
  {title} {Respiratory aerosols and droplets in the transmission of infectious
  diseases},\ }\href {https://doi.org/10.1103/RevModPhys.95.045001} {\bibfield
  {journal} {\bibinfo  {journal} {Rev. Mod. Phys.}\ }\textbf {\bibinfo {volume}
  {95}},\ \bibinfo {pages} {045001} (\bibinfo {year} {2023})}\BibitemShut
  {NoStop}%
\bibitem [{\citenamefont {Abkarian}\ \emph {et~al.}(2020)\citenamefont
  {Abkarian}, \citenamefont {Mendez}, \citenamefont {Xue}, \citenamefont
  {Yang},\ and\ \citenamefont {Stone}}]{abkarian2020}%
  \BibitemOpen
  \bibfield  {author} {\bibinfo {author} {\bibfnamefont {M.}~\bibnamefont
  {Abkarian}}, \bibinfo {author} {\bibfnamefont {S.}~\bibnamefont {Mendez}},
  \bibinfo {author} {\bibfnamefont {N.}~\bibnamefont {Xue}}, \bibinfo {author}
  {\bibfnamefont {F.}~\bibnamefont {Yang}},\ and\ \bibinfo {author}
  {\bibfnamefont {H.~A.}\ \bibnamefont {Stone}},\ }\bibfield  {title} {\bibinfo
  {title} {Speech can produce jet-like transport relevant to asymptomatic
  spreading of virus},\ }\href
  {https://doi.org/https://doi.org/10.1073/pnas.2012156117} {\bibfield
  {journal} {\bibinfo  {journal} {Proceedings of the National Academy of
  Sciences}\ }\textbf {\bibinfo {volume} {117}},\ \bibinfo {pages} {25237}
  (\bibinfo {year} {2020})}\BibitemShut {NoStop}%
\bibitem [{\citenamefont {Adenaiye}\ \emph {et~al.}(2021)\citenamefont
  {Adenaiye}, \citenamefont {Lai}, \citenamefont {Bueno~de Mesquita},
  \citenamefont {Hong}, \citenamefont {Youssefi}, \citenamefont {German},
  \citenamefont {Tai}, \citenamefont {Albert}, \citenamefont {Schanz},
  \citenamefont {Weston}, \citenamefont {Hang}, \citenamefont {Fung},
  \citenamefont {Chung}, \citenamefont {Coleman}, \citenamefont {Sapoval},
  \citenamefont {Treangen}, \citenamefont {Berry}, \citenamefont {Mullins},
  \citenamefont {Frieman}, \citenamefont {Ma},\ and\ \citenamefont
  {Milton}}]{adenaiye2021}%
  \BibitemOpen
  \bibfield  {author} {\bibinfo {author} {\bibfnamefont {O.~O.}\ \bibnamefont
  {Adenaiye}}, \bibinfo {author} {\bibfnamefont {J.}~\bibnamefont {Lai}},
  \bibinfo {author} {\bibfnamefont {P.~J.}\ \bibnamefont {Bueno~de Mesquita}},
  \bibinfo {author} {\bibfnamefont {F.}~\bibnamefont {Hong}}, \bibinfo {author}
  {\bibfnamefont {S.}~\bibnamefont {Youssefi}}, \bibinfo {author}
  {\bibfnamefont {J.}~\bibnamefont {German}}, \bibinfo {author} {\bibfnamefont
  {S.~H.~S.}\ \bibnamefont {Tai}}, \bibinfo {author} {\bibfnamefont
  {B.}~\bibnamefont {Albert}}, \bibinfo {author} {\bibfnamefont
  {M.}~\bibnamefont {Schanz}}, \bibinfo {author} {\bibfnamefont
  {S.}~\bibnamefont {Weston}}, \bibinfo {author} {\bibfnamefont
  {J.}~\bibnamefont {Hang}}, \bibinfo {author} {\bibfnamefont {C.}~\bibnamefont
  {Fung}}, \bibinfo {author} {\bibfnamefont {H.~K.}\ \bibnamefont {Chung}},
  \bibinfo {author} {\bibfnamefont {K.~K.}\ \bibnamefont {Coleman}}, \bibinfo
  {author} {\bibfnamefont {N.}~\bibnamefont {Sapoval}}, \bibinfo {author}
  {\bibfnamefont {T.}~\bibnamefont {Treangen}}, \bibinfo {author}
  {\bibfnamefont {I.~M.}\ \bibnamefont {Berry}}, \bibinfo {author}
  {\bibfnamefont {K.}~\bibnamefont {Mullins}}, \bibinfo {author} {\bibfnamefont
  {M.}~\bibnamefont {Frieman}}, \bibinfo {author} {\bibfnamefont
  {T.}~\bibnamefont {Ma}},\ and\ \bibinfo {author} {\bibfnamefont {U.~o. M. S.
  R.~G.}\ \bibnamefont {Milton}, \bibfnamefont {Donald~K}},\ }\bibfield
  {title} {\bibinfo {title} {{Infectious Severe Acute Respiratory Syndrome
  Coronavirus 2 (SARS-CoV-2) in Exhaled Aerosols and Efficacy of Masks During
  Early Mild Infection}},\ }\href {https://doi.org/10.1093/cid/ciab797}
  {\bibfield  {journal} {\bibinfo  {journal} {Clinical Infectious Diseases}\
  }\textbf {\bibinfo {volume} {75}},\ \bibinfo {pages} {e241} (\bibinfo {year}
  {2021})}\BibitemShut {NoStop}%
\bibitem [{\citenamefont {Tan}\ \emph {et~al.}(2023)\citenamefont {Tan},
  \citenamefont {Ong}, \citenamefont {Koh}, \citenamefont {Tay}, \citenamefont
  {Aw}, \citenamefont {Nah}, \citenamefont {Abdullah}, \citenamefont {Coleman},
  \citenamefont {Milton}, \citenamefont {Chu}, \citenamefont {Chow},
  \citenamefont {Tambyah},\ and\ \citenamefont {Tham}}]{tan2023}%
  \BibitemOpen
  \bibfield  {author} {\bibinfo {author} {\bibfnamefont {K.~S.}\ \bibnamefont
  {Tan}}, \bibinfo {author} {\bibfnamefont {S.~W.~X.}\ \bibnamefont {Ong}},
  \bibinfo {author} {\bibfnamefont {M.~H.}\ \bibnamefont {Koh}}, \bibinfo
  {author} {\bibfnamefont {D.~J.~W.}\ \bibnamefont {Tay}}, \bibinfo {author}
  {\bibfnamefont {D.~Z.~H.}\ \bibnamefont {Aw}}, \bibinfo {author}
  {\bibfnamefont {Y.~W.}\ \bibnamefont {Nah}}, \bibinfo {author} {\bibfnamefont
  {M.~R.~B.}\ \bibnamefont {Abdullah}}, \bibinfo {author} {\bibfnamefont
  {K.~K.}\ \bibnamefont {Coleman}}, \bibinfo {author} {\bibfnamefont {D.~K.}\
  \bibnamefont {Milton}}, \bibinfo {author} {\bibfnamefont {J.~J.~H.}\
  \bibnamefont {Chu}}, \bibinfo {author} {\bibfnamefont {V.~T.}\ \bibnamefont
  {Chow}}, \bibinfo {author} {\bibfnamefont {P.~A.}\ \bibnamefont {Tambyah}},\
  and\ \bibinfo {author} {\bibfnamefont {K.~W.}\ \bibnamefont {Tham}},\
  }\bibfield  {title} {\bibinfo {title} {{SARS-CoV-2 Omicron variant shedding
  during respiratory activities}},\ }\href
  {https://doi.org/https://doi.org/10.1016/j.ijid.2023.03.029} {\bibfield
  {journal} {\bibinfo  {journal} {International Journal of Infectious
  Diseases}\ }\textbf {\bibinfo {volume} {131}},\ \bibinfo {pages} {19}
  (\bibinfo {year} {2023})}\BibitemShut {NoStop}%
\bibitem [{\citenamefont {Jia}\ \emph {et~al.}(2022)\citenamefont {Jia},
  \citenamefont {Wei}, \citenamefont {Cheng}, \citenamefont {Wang},\ and\
  \citenamefont {Li}}]{jia2022}%
  \BibitemOpen
  \bibfield  {author} {\bibinfo {author} {\bibfnamefont {W.}~\bibnamefont
  {Jia}}, \bibinfo {author} {\bibfnamefont {J.}~\bibnamefont {Wei}}, \bibinfo
  {author} {\bibfnamefont {P.}~\bibnamefont {Cheng}}, \bibinfo {author}
  {\bibfnamefont {Q.}~\bibnamefont {Wang}},\ and\ \bibinfo {author}
  {\bibfnamefont {Y.}~\bibnamefont {Li}},\ }\bibfield  {title} {\bibinfo
  {title} {Exposure and respiratory infection risk via the short-range airborne
  route},\ }\href
  {https://doi.org/https://doi.org/10.1016/j.buildenv.2022.109166} {\bibfield
  {journal} {\bibinfo  {journal} {Building and Environment}\ }\textbf {\bibinfo
  {volume} {219}},\ \bibinfo {pages} {109166} (\bibinfo {year}
  {2022})}\BibitemShut {NoStop}%
\bibitem [{\citenamefont {Archer}\ \emph {et~al.}(2022)\citenamefont {Archer},
  \citenamefont {McCarthy}, \citenamefont {Symons}, \citenamefont {Watson},
  \citenamefont {Orton}, \citenamefont {Browne}, \citenamefont {Harrison},
  \citenamefont {Moseley}, \citenamefont {Philip}, \citenamefont {Calder} \emph
  {et~al.}}]{archer2022}%
  \BibitemOpen
  \bibfield  {author} {\bibinfo {author} {\bibfnamefont {J.}~\bibnamefont
  {Archer}}, \bibinfo {author} {\bibfnamefont {L.~P.}\ \bibnamefont
  {McCarthy}}, \bibinfo {author} {\bibfnamefont {H.~E.}\ \bibnamefont
  {Symons}}, \bibinfo {author} {\bibfnamefont {N.~A.}\ \bibnamefont {Watson}},
  \bibinfo {author} {\bibfnamefont {C.~M.}\ \bibnamefont {Orton}}, \bibinfo
  {author} {\bibfnamefont {W.~J.}\ \bibnamefont {Browne}}, \bibinfo {author}
  {\bibfnamefont {J.}~\bibnamefont {Harrison}}, \bibinfo {author}
  {\bibfnamefont {B.}~\bibnamefont {Moseley}}, \bibinfo {author} {\bibfnamefont
  {K.~E.}\ \bibnamefont {Philip}}, \bibinfo {author} {\bibfnamefont {J.~D.}\
  \bibnamefont {Calder}}, \emph {et~al.},\ }\bibfield  {title} {\bibinfo
  {title} {Comparing aerosol number and mass exhalation rates from children and
  adults during breathing, speaking and singing},\ }\href
  {https://doi.org/https://doi.org/10.1098/rsfs.2021.0078} {\bibfield
  {journal} {\bibinfo  {journal} {Interface Focus}\ }\textbf {\bibinfo {volume}
  {12}},\ \bibinfo {pages} {20210078} (\bibinfo {year} {2022})}\BibitemShut
  {NoStop}%
\bibitem [{\citenamefont {Jones}\ \emph {et~al.}(2021)\citenamefont {Jones},
  \citenamefont {Biele}, \citenamefont {Mühlemann}, \citenamefont {Veith},
  \citenamefont {Schneider}, \citenamefont {Beheim-Schwarzbach}, \citenamefont
  {Bleicker}, \citenamefont {Tesch}, \citenamefont {Schmidt}, \citenamefont
  {Sander}, \citenamefont {Kurth}, \citenamefont {Menzel}, \citenamefont
  {Schwarzer}, \citenamefont {Zuchowski}, \citenamefont {Hofmann},
  \citenamefont {Krumbholz}, \citenamefont {Stein}, \citenamefont {Edelmann},
  \citenamefont {Corman},\ and\ \citenamefont {Drosten}}]{jones2021}%
  \BibitemOpen
  \bibfield  {author} {\bibinfo {author} {\bibfnamefont {T.~C.}\ \bibnamefont
  {Jones}}, \bibinfo {author} {\bibfnamefont {G.}~\bibnamefont {Biele}},
  \bibinfo {author} {\bibfnamefont {B.}~\bibnamefont {Mühlemann}}, \bibinfo
  {author} {\bibfnamefont {T.}~\bibnamefont {Veith}}, \bibinfo {author}
  {\bibfnamefont {J.}~\bibnamefont {Schneider}}, \bibinfo {author}
  {\bibfnamefont {J.}~\bibnamefont {Beheim-Schwarzbach}}, \bibinfo {author}
  {\bibfnamefont {T.}~\bibnamefont {Bleicker}}, \bibinfo {author}
  {\bibfnamefont {J.}~\bibnamefont {Tesch}}, \bibinfo {author} {\bibfnamefont
  {M.~L.}\ \bibnamefont {Schmidt}}, \bibinfo {author} {\bibfnamefont {L.~E.}\
  \bibnamefont {Sander}}, \bibinfo {author} {\bibfnamefont {F.}~\bibnamefont
  {Kurth}}, \bibinfo {author} {\bibfnamefont {P.}~\bibnamefont {Menzel}},
  \bibinfo {author} {\bibfnamefont {R.}~\bibnamefont {Schwarzer}}, \bibinfo
  {author} {\bibfnamefont {M.}~\bibnamefont {Zuchowski}}, \bibinfo {author}
  {\bibfnamefont {J.}~\bibnamefont {Hofmann}}, \bibinfo {author} {\bibfnamefont
  {A.}~\bibnamefont {Krumbholz}}, \bibinfo {author} {\bibfnamefont
  {A.}~\bibnamefont {Stein}}, \bibinfo {author} {\bibfnamefont
  {A.}~\bibnamefont {Edelmann}}, \bibinfo {author} {\bibfnamefont {V.~M.}\
  \bibnamefont {Corman}},\ and\ \bibinfo {author} {\bibfnamefont
  {C.}~\bibnamefont {Drosten}},\ }\bibfield  {title} {\bibinfo {title}
  {{Estimating infectiousness throughout SARS-CoV-2 infection course}},\ }\href
  {https://doi.org/10.1126/science.abi5273} {\bibfield  {journal} {\bibinfo
  {journal} {Science}\ }\textbf {\bibinfo {volume} {373}},\ \bibinfo {pages}
  {eabi5273} (\bibinfo {year} {2021})}\BibitemShut {NoStop}%
\bibitem [{\citenamefont {Greenhalgh}\ \emph {et~al.}(2024)\citenamefont
  {Greenhalgh}, \citenamefont {MacIntyre}, \citenamefont {Baker}, \citenamefont
  {Bhattacharjee}, \citenamefont {Chughtai}, \citenamefont {Fisman},
  \citenamefont {Kunasekaran}, \citenamefont {Kvalsvig}, \citenamefont
  {Lupton}, \citenamefont {Oliver}, \citenamefont {Tawfiq}, \citenamefont
  {Ungrin},\ and\ \citenamefont {Vipond}}]{greenhalgh2024}%
  \BibitemOpen
  \bibfield  {author} {\bibinfo {author} {\bibfnamefont {T.}~\bibnamefont
  {Greenhalgh}}, \bibinfo {author} {\bibfnamefont {C.~R.}\ \bibnamefont
  {MacIntyre}}, \bibinfo {author} {\bibfnamefont {M.~G.}\ \bibnamefont
  {Baker}}, \bibinfo {author} {\bibfnamefont {S.}~\bibnamefont
  {Bhattacharjee}}, \bibinfo {author} {\bibfnamefont {A.~A.}\ \bibnamefont
  {Chughtai}}, \bibinfo {author} {\bibfnamefont {D.}~\bibnamefont {Fisman}},
  \bibinfo {author} {\bibfnamefont {M.}~\bibnamefont {Kunasekaran}}, \bibinfo
  {author} {\bibfnamefont {A.}~\bibnamefont {Kvalsvig}}, \bibinfo {author}
  {\bibfnamefont {D.}~\bibnamefont {Lupton}}, \bibinfo {author} {\bibfnamefont
  {M.}~\bibnamefont {Oliver}}, \bibinfo {author} {\bibfnamefont
  {E.}~\bibnamefont {Tawfiq}}, \bibinfo {author} {\bibfnamefont
  {M.}~\bibnamefont {Ungrin}},\ and\ \bibinfo {author} {\bibfnamefont
  {J.}~\bibnamefont {Vipond}},\ }\bibfield  {title} {\bibinfo {title} {Masks
  and respirators for prevention of respiratory infections: a state of the
  science review},\ }\href {https://doi.org/10.1128/cmr.00124-23} {\bibfield
  {journal} {\bibinfo  {journal} {Clinical Microbiology Reviews}\ }\textbf
  {\bibinfo {volume} {37}},\ \bibinfo {pages} {e00124} (\bibinfo {year}
  {2024})}\BibitemShut {NoStop}%
\bibitem [{\citenamefont {Bagheri}\ \emph {et~al.}(2021)\citenamefont
  {Bagheri}, \citenamefont {Thiede}, \citenamefont {Hejazi}, \citenamefont
  {Schlenczek},\ and\ \citenamefont {Bodenschatz}}]{bagheri2021}%
  \BibitemOpen
  \bibfield  {author} {\bibinfo {author} {\bibfnamefont {G.}~\bibnamefont
  {Bagheri}}, \bibinfo {author} {\bibfnamefont {B.}~\bibnamefont {Thiede}},
  \bibinfo {author} {\bibfnamefont {B.}~\bibnamefont {Hejazi}}, \bibinfo
  {author} {\bibfnamefont {O.}~\bibnamefont {Schlenczek}},\ and\ \bibinfo
  {author} {\bibfnamefont {E.}~\bibnamefont {Bodenschatz}},\ }\bibfield
  {title} {\bibinfo {title} {An upper bound on one-to-one exposure to
  infectious human respiratory particles},\ }\href
  {https://doi.org/https://doi.org/10.1073/pnas.2110117118} {\bibfield
  {journal} {\bibinfo  {journal} {Proceedings of the National Academy of
  Sciences}\ }\textbf {\bibinfo {volume} {118}},\ \bibinfo {pages}
  {e2110117118} (\bibinfo {year} {2021})}\BibitemShut {NoStop}%
\bibitem [{\citenamefont {Robinson}\ \emph {et~al.}(2022)\citenamefont
  {Robinson}, \citenamefont {{Rios de Anda}}, \citenamefont {Moore},
  \citenamefont {Gregson}, \citenamefont {Reid}, \citenamefont {Husain},
  \citenamefont {Sear},\ and\ \citenamefont {Royall}}]{robinson2022}%
  \BibitemOpen
  \bibfield  {author} {\bibinfo {author} {\bibfnamefont {J.~F.}\ \bibnamefont
  {Robinson}}, \bibinfo {author} {\bibfnamefont {I.}~\bibnamefont {{Rios de
  Anda}}}, \bibinfo {author} {\bibfnamefont {F.~J.}\ \bibnamefont {Moore}},
  \bibinfo {author} {\bibfnamefont {F.~K.~A.}\ \bibnamefont {Gregson}},
  \bibinfo {author} {\bibfnamefont {J.~P.}\ \bibnamefont {Reid}}, \bibinfo
  {author} {\bibfnamefont {L.}~\bibnamefont {Husain}}, \bibinfo {author}
  {\bibfnamefont {R.~P.}\ \bibnamefont {Sear}},\ and\ \bibinfo {author}
  {\bibfnamefont {C.~P.}\ \bibnamefont {Royall}},\ }\bibfield  {title}
  {\bibinfo {title} {How effective are face coverings in reducing transmission
  of {{COVID-19}}?},\ }\href {https://doi.org/10.1080/02786826.2022.2042467}
  {\bibfield  {journal} {\bibinfo  {journal} {Aerosol Science Technology}\
  }\textbf {\bibinfo {volume} {56}},\ \bibinfo {pages} {473} (\bibinfo {year}
  {2022})}\BibitemShut {NoStop}%
\bibitem [{\citenamefont {Zoller}\ \emph {et~al.}(2021)\citenamefont {Zoller},
  \citenamefont {Meyer},\ and\ \citenamefont {Dittler}}]{zoller2021}%
  \BibitemOpen
  \bibfield  {author} {\bibinfo {author} {\bibfnamefont {J.}~\bibnamefont
  {Zoller}}, \bibinfo {author} {\bibfnamefont {J.}~\bibnamefont {Meyer}},\ and\
  \bibinfo {author} {\bibfnamefont {A.}~\bibnamefont {Dittler}},\ }\bibfield
  {title} {\bibinfo {title} {{A critical note on filtering-face-piece
  filtration efficiency determination applying EN 149}},\ }\href
  {https://doi.org/https://doi.org/10.1016/j.jaerosci.2021.105830} {\bibfield
  {journal} {\bibinfo  {journal} {J. Aerosol Science}\ }\textbf {\bibinfo
  {volume} {158}},\ \bibinfo {pages} {105830} (\bibinfo {year}
  {2021})}\BibitemShut {NoStop}%
\bibitem [{\citenamefont {Duncan}\ \emph {et~al.}(2021)\citenamefont {Duncan},
  \citenamefont {Bodurtha},\ and\ \citenamefont {Naqvi}}]{duncan2021}%
  \BibitemOpen
  \bibfield  {author} {\bibinfo {author} {\bibfnamefont {S.}~\bibnamefont
  {Duncan}}, \bibinfo {author} {\bibfnamefont {P.}~\bibnamefont {Bodurtha}},\
  and\ \bibinfo {author} {\bibfnamefont {S.}~\bibnamefont {Naqvi}},\ }\bibfield
   {title} {\bibinfo {title} {N95 respirators, disposable procedure masks and
  reusable cloth face coverings: Total inward leakage and filtration efficiency
  of materials against aerosol},\ }\href
  {https://doi.org/https://doi.org/10.1371/journal.pone.0258191} {\bibfield
  {journal} {\bibinfo  {journal} {PloS ONE}\ }\textbf {\bibinfo {volume}
  {16}},\ \bibinfo {pages} {e0258191} (\bibinfo {year} {2021})}\BibitemShut
  {NoStop}%
\bibitem [{\citenamefont {Johnston}(2006)}]{johnston2006}%
  \BibitemOpen
  \bibfield  {author} {\bibinfo {author} {\bibfnamefont {D.}~\bibnamefont
  {Johnston}},\ }\bibfield  {title} {\bibinfo {title} {Stretched exponential
  relaxation arising from a continuous sum of exponential decays},\ }\href
  {https://doi.org/https://doi.org/10.1103/PhysRevB.74.184430} {\bibfield
  {journal} {\bibinfo  {journal} {Physical Review B}\ }\textbf {\bibinfo
  {volume} {74}},\ \bibinfo {pages} {184430} (\bibinfo {year}
  {2006})}\BibitemShut {NoStop}%
\bibitem [{\citenamefont {Newman}(2005)}]{newman2005}%
  \BibitemOpen
  \bibfield  {author} {\bibinfo {author} {\bibfnamefont {M.~E.}\ \bibnamefont
  {Newman}},\ }\bibfield  {title} {\bibinfo {title} {{Power laws, Pareto
  distributions and Zipf's law}},\ }\href
  {https://doi.org/https://doi.org/10.1080/00107510500052444} {\bibfield
  {journal} {\bibinfo  {journal} {Contemporary physics}\ }\textbf {\bibinfo
  {volume} {46}},\ \bibinfo {pages} {323} (\bibinfo {year} {2005})}\BibitemShut
  {NoStop}%
\bibitem [{\citenamefont {Clauset}\ \emph {et~al.}(2009)\citenamefont
  {Clauset}, \citenamefont {Shalizi},\ and\ \citenamefont
  {Newman}}]{clauset2009}%
  \BibitemOpen
  \bibfield  {author} {\bibinfo {author} {\bibfnamefont {A.}~\bibnamefont
  {Clauset}}, \bibinfo {author} {\bibfnamefont {C.~R.}\ \bibnamefont
  {Shalizi}},\ and\ \bibinfo {author} {\bibfnamefont {M.~E.}\ \bibnamefont
  {Newman}},\ }\bibfield  {title} {\bibinfo {title} {Power-law distributions in
  empirical data},\ }\href {https://doi.org/https://doi.org/10.1137/070710111}
  {\bibfield  {journal} {\bibinfo  {journal} {SIAM review}\ }\textbf {\bibinfo
  {volume} {51}},\ \bibinfo {pages} {661} (\bibinfo {year} {2009})}\BibitemShut
  {NoStop}%
\bibitem [{\citenamefont {Lindsey}\ and\ \citenamefont
  {Patterson}(1980)}]{lindsey1980}%
  \BibitemOpen
  \bibfield  {author} {\bibinfo {author} {\bibfnamefont {C.~P.}\ \bibnamefont
  {Lindsey}}\ and\ \bibinfo {author} {\bibfnamefont {G.~D.}\ \bibnamefont
  {Patterson}},\ }\bibfield  {title} {\bibinfo {title} {{Detailed comparison of
  the Williams–Watts and Cole–Davidson functions}},\ }\href
  {https://doi.org/10.1063/1.440530} {\bibfield  {journal} {\bibinfo  {journal}
  {Journal Chemical Physics}\ }\textbf {\bibinfo {volume} {73}},\ \bibinfo
  {pages} {3348} (\bibinfo {year} {1980})}\BibitemShut {NoStop}%
\bibitem [{\citenamefont {Wu}\ and\ \citenamefont {Jia}(2016)}]{wu2016}%
  \BibitemOpen
  \bibfield  {author} {\bibinfo {author} {\bibfnamefont {J.}~\bibnamefont
  {Wu}}\ and\ \bibinfo {author} {\bibfnamefont {Q.}~\bibnamefont {Jia}},\
  }\bibfield  {title} {\bibinfo {title} {{The heterogeneous energy landscape
  expression of KWW relaxation}},\ }\href
  {https://doi.org/https://doi.org/10.1038/srep20506} {\bibfield  {journal}
  {\bibinfo  {journal} {Scientific reports}\ }\textbf {\bibinfo {volume} {6}},\
  \bibinfo {pages} {20506} (\bibinfo {year} {2016})}\BibitemShut {NoStop}%
\bibitem [{\citenamefont {Kendall}\ \emph {et~al.}(2023)\citenamefont
  {Kendall}, \citenamefont {Tsallis}, \citenamefont {Wymant}, \citenamefont
  {Di~Francia}, \citenamefont {Balogun}, \citenamefont {Didelot}, \citenamefont
  {Ferretti},\ and\ \citenamefont {Fraser}}]{kendall2023}%
  \BibitemOpen
  \bibfield  {author} {\bibinfo {author} {\bibfnamefont {M.}~\bibnamefont
  {Kendall}}, \bibinfo {author} {\bibfnamefont {D.}~\bibnamefont {Tsallis}},
  \bibinfo {author} {\bibfnamefont {C.}~\bibnamefont {Wymant}}, \bibinfo
  {author} {\bibfnamefont {A.}~\bibnamefont {Di~Francia}}, \bibinfo {author}
  {\bibfnamefont {Y.}~\bibnamefont {Balogun}}, \bibinfo {author} {\bibfnamefont
  {X.}~\bibnamefont {Didelot}}, \bibinfo {author} {\bibfnamefont
  {L.}~\bibnamefont {Ferretti}},\ and\ \bibinfo {author} {\bibfnamefont
  {C.}~\bibnamefont {Fraser}},\ }\bibfield  {title} {\bibinfo {title}
  {{Epidemiological impacts of the NHS COVID-19 app in England and Wales
  throughout its first year}},\ }\href
  {https://doi.org/https://doi.org/10.1038/s41467-023-36495-z} {\bibfield
  {journal} {\bibinfo  {journal} {Nature Communications}\ }\textbf {\bibinfo
  {volume} {14}},\ \bibinfo {pages} {858} (\bibinfo {year} {2023})}\BibitemShut
  {NoStop}%
\bibitem [{\citenamefont {Ferretti}\ \emph {et~al.}(2024)\citenamefont
  {Ferretti}, \citenamefont {Wymant}, \citenamefont {Petrie}, \citenamefont
  {Tsallis}, \citenamefont {Kendall}, \citenamefont {Ledda}, \citenamefont
  {Di~Lauro}, \citenamefont {Fowler}, \citenamefont {Di~Francia}, \citenamefont
  {Panovska-Griffiths} \emph {et~al.}}]{ferretti2024}%
  \BibitemOpen
  \bibfield  {author} {\bibinfo {author} {\bibfnamefont {L.}~\bibnamefont
  {Ferretti}}, \bibinfo {author} {\bibfnamefont {C.}~\bibnamefont {Wymant}},
  \bibinfo {author} {\bibfnamefont {J.}~\bibnamefont {Petrie}}, \bibinfo
  {author} {\bibfnamefont {D.}~\bibnamefont {Tsallis}}, \bibinfo {author}
  {\bibfnamefont {M.}~\bibnamefont {Kendall}}, \bibinfo {author} {\bibfnamefont
  {A.}~\bibnamefont {Ledda}}, \bibinfo {author} {\bibfnamefont
  {F.}~\bibnamefont {Di~Lauro}}, \bibinfo {author} {\bibfnamefont
  {A.}~\bibnamefont {Fowler}}, \bibinfo {author} {\bibfnamefont
  {A.}~\bibnamefont {Di~Francia}}, \bibinfo {author} {\bibfnamefont
  {J.}~\bibnamefont {Panovska-Griffiths}}, \emph {et~al.},\ }\bibfield  {title}
  {\bibinfo {title} {{Digital measurement of SARS-CoV-2 transmission risk from
  7 million contacts}},\ }\href
  {https://doi.org/https://doi.org/10.1038/s41586-023-06952-2} {\bibfield
  {journal} {\bibinfo  {journal} {Nature}\ ,\ \bibinfo {pages} {145–150}}
  (\bibinfo {year} {2024})}\BibitemShut {NoStop}%
\bibitem [{\citenamefont {Takatsuki}\ \emph {et~al.}(2023)\citenamefont
  {Takatsuki}, \citenamefont {Takahashi}, \citenamefont {Nakajima},
  \citenamefont {Iwasaki}, \citenamefont {Nagano}, \citenamefont {Tani-Sassa},
  \citenamefont {Yuasa}, \citenamefont {Kanehira}, \citenamefont {Sonobe},
  \citenamefont {Nukui}, \citenamefont {Takeuchi}, \citenamefont {Tanimoto},
  \citenamefont {Tanaka}, \citenamefont {Kimura}, \citenamefont {Ichimura},\
  and\ \citenamefont {Tohda}}]{takatsuki2023}%
  \BibitemOpen
  \bibfield  {author} {\bibinfo {author} {\bibfnamefont {Y.}~\bibnamefont
  {Takatsuki}}, \bibinfo {author} {\bibfnamefont {Y.}~\bibnamefont
  {Takahashi}}, \bibinfo {author} {\bibfnamefont {J.}~\bibnamefont {Nakajima}},
  \bibinfo {author} {\bibfnamefont {Y.}~\bibnamefont {Iwasaki}}, \bibinfo
  {author} {\bibfnamefont {K.}~\bibnamefont {Nagano}}, \bibinfo {author}
  {\bibfnamefont {C.}~\bibnamefont {Tani-Sassa}}, \bibinfo {author}
  {\bibfnamefont {S.}~\bibnamefont {Yuasa}}, \bibinfo {author} {\bibfnamefont
  {S.}~\bibnamefont {Kanehira}}, \bibinfo {author} {\bibfnamefont
  {K.}~\bibnamefont {Sonobe}}, \bibinfo {author} {\bibfnamefont
  {Y.}~\bibnamefont {Nukui}}, \bibinfo {author} {\bibfnamefont
  {H.}~\bibnamefont {Takeuchi}}, \bibinfo {author} {\bibfnamefont
  {K.}~\bibnamefont {Tanimoto}}, \bibinfo {author} {\bibfnamefont
  {Y.}~\bibnamefont {Tanaka}}, \bibinfo {author} {\bibfnamefont
  {A.}~\bibnamefont {Kimura}}, \bibinfo {author} {\bibfnamefont
  {N.}~\bibnamefont {Ichimura}},\ and\ \bibinfo {author} {\bibfnamefont
  {S.}~\bibnamefont {Tohda}},\ }\bibfield  {title} {\bibinfo {title} {Viral
  load of sars-cov-2 omicron ba.5 is lower than that of ba.2 despite the higher
  infectivity of ba.5},\ }\href
  {https://doi.org/https://doi.org/10.1002/iid3.783} {\bibfield  {journal}
  {\bibinfo  {journal} {Immunity, Inflammation and Disease}\ }\textbf {\bibinfo
  {volume} {11}},\ \bibinfo {pages} {e783} (\bibinfo {year}
  {2023})}\BibitemShut {NoStop}%
\bibitem [{\citenamefont {Peng}\ \emph {et~al.}(2022)\citenamefont {Peng},
  \citenamefont {Rojas}, \citenamefont {Kropff}, \citenamefont {Bahnfleth},
  \citenamefont {Buonanno}, \citenamefont {Dancer}, \citenamefont {Kurnitski},
  \citenamefont {Li}, \citenamefont {Loomans}, \citenamefont {Marr},
  \citenamefont {Morawska}, \citenamefont {Nazaroff}, \citenamefont {Noakes},
  \citenamefont {Querol}, \citenamefont {Sekhar}, \citenamefont {Tellier},
  \citenamefont {Greenhalgh}, \citenamefont {Bourouiba}, \citenamefont
  {Boerstra}, \citenamefont {Tang}, \citenamefont {Miller},\ and\ \citenamefont
  {Jimenez}}]{peng2022}%
  \BibitemOpen
  \bibfield  {author} {\bibinfo {author} {\bibfnamefont {Z.}~\bibnamefont
  {Peng}}, \bibinfo {author} {\bibfnamefont {A.~P.}\ \bibnamefont {Rojas}},
  \bibinfo {author} {\bibfnamefont {E.}~\bibnamefont {Kropff}}, \bibinfo
  {author} {\bibfnamefont {W.}~\bibnamefont {Bahnfleth}}, \bibinfo {author}
  {\bibfnamefont {G.}~\bibnamefont {Buonanno}}, \bibinfo {author}
  {\bibfnamefont {S.}~\bibnamefont {Dancer}}, \bibinfo {author} {\bibfnamefont
  {J.}~\bibnamefont {Kurnitski}}, \bibinfo {author} {\bibfnamefont
  {Y.}~\bibnamefont {Li}}, \bibinfo {author} {\bibfnamefont {M.}~\bibnamefont
  {Loomans}}, \bibinfo {author} {\bibfnamefont {L.}~\bibnamefont {Marr}},
  \bibinfo {author} {\bibfnamefont {L.}~\bibnamefont {Morawska}}, \bibinfo
  {author} {\bibfnamefont {W.}~\bibnamefont {Nazaroff}}, \bibinfo {author}
  {\bibfnamefont {C.}~\bibnamefont {Noakes}}, \bibinfo {author} {\bibfnamefont
  {X.}~\bibnamefont {Querol}}, \bibinfo {author} {\bibfnamefont
  {C.}~\bibnamefont {Sekhar}}, \bibinfo {author} {\bibfnamefont
  {R.}~\bibnamefont {Tellier}}, \bibinfo {author} {\bibfnamefont
  {T.}~\bibnamefont {Greenhalgh}}, \bibinfo {author} {\bibfnamefont
  {L.}~\bibnamefont {Bourouiba}}, \bibinfo {author} {\bibfnamefont
  {A.}~\bibnamefont {Boerstra}}, \bibinfo {author} {\bibfnamefont
  {J.}~\bibnamefont {Tang}}, \bibinfo {author} {\bibfnamefont {S.}~\bibnamefont
  {Miller}},\ and\ \bibinfo {author} {\bibfnamefont {J.}~\bibnamefont
  {Jimenez}},\ }\bibfield  {title} {\bibinfo {title} {Practical indicators for
  risk of airborne transmission in shared indoor environments and their
  application to covid-19 outbreaks},\ }\href
  {https://doi.org/10.1021/acs.est.1c06531} {\bibfield  {journal} {\bibinfo
  {journal} {Environmental Science \& Technology}\ }\textbf {\bibinfo {volume}
  {56}},\ \bibinfo {pages} {1125} (\bibinfo {year} {2022})}\BibitemShut
  {NoStop}%
\bibitem [{\citenamefont {Puhach}\ \emph {et~al.}(2023)\citenamefont {Puhach},
  \citenamefont {Meyer},\ and\ \citenamefont {Eckerle}}]{puhach2023}%
  \BibitemOpen
  \bibfield  {author} {\bibinfo {author} {\bibfnamefont {O.}~\bibnamefont
  {Puhach}}, \bibinfo {author} {\bibfnamefont {B.}~\bibnamefont {Meyer}},\ and\
  \bibinfo {author} {\bibfnamefont {I.}~\bibnamefont {Eckerle}},\ }\bibfield
  {title} {\bibinfo {title} {{SARS-CoV-2 viral load and shedding kinetics}},\
  }\href {https://doi.org/https://doi.org/10.1038/s41579-022-00822-w}
  {\bibfield  {journal} {\bibinfo  {journal} {Nature Reviews Microbiology}\
  }\textbf {\bibinfo {volume} {21}},\ \bibinfo {pages} {147} (\bibinfo {year}
  {2023})}\BibitemShut {NoStop}%
\bibitem [{\citenamefont {Killingley}\ \emph {et~al.}(2022)\citenamefont
  {Killingley}, \citenamefont {Mann}, \citenamefont {Kalinova}, \citenamefont
  {Boyers}, \citenamefont {Goonawardane}, \citenamefont {Zhou}, \citenamefont
  {Lindsell}, \citenamefont {Hare}, \citenamefont {Brown}, \citenamefont
  {Frise} \emph {et~al.}}]{killingley2022}%
  \BibitemOpen
  \bibfield  {author} {\bibinfo {author} {\bibfnamefont {B.}~\bibnamefont
  {Killingley}}, \bibinfo {author} {\bibfnamefont {A.~J.}\ \bibnamefont
  {Mann}}, \bibinfo {author} {\bibfnamefont {M.}~\bibnamefont {Kalinova}},
  \bibinfo {author} {\bibfnamefont {A.}~\bibnamefont {Boyers}}, \bibinfo
  {author} {\bibfnamefont {N.}~\bibnamefont {Goonawardane}}, \bibinfo {author}
  {\bibfnamefont {J.}~\bibnamefont {Zhou}}, \bibinfo {author} {\bibfnamefont
  {K.}~\bibnamefont {Lindsell}}, \bibinfo {author} {\bibfnamefont {S.~S.}\
  \bibnamefont {Hare}}, \bibinfo {author} {\bibfnamefont {J.}~\bibnamefont
  {Brown}}, \bibinfo {author} {\bibfnamefont {R.}~\bibnamefont {Frise}}, \emph
  {et~al.},\ }\bibfield  {title} {\bibinfo {title} {{Safety, tolerability and
  viral kinetics during SARS-CoV-2 human challenge in young adults}},\ }\href
  {https://doi.org/https://doi.org/10.1038/s41591-022-01780-9} {\bibfield
  {journal} {\bibinfo  {journal} {Nature Medicine}\ }\textbf {\bibinfo {volume}
  {28}},\ \bibinfo {pages} {1031} (\bibinfo {year} {2022})}\BibitemShut
  {NoStop}%
\bibitem [{\citenamefont {Giri}\ \emph {et~al.}(2022)\citenamefont {Giri},
  \citenamefont {Biswas}, \citenamefont {Chase}, \citenamefont {Xue},
  \citenamefont {Abkarian}, \citenamefont {Mendez}, \citenamefont {Saha},\ and\
  \citenamefont {Stone}}]{giri2022}%
  \BibitemOpen
  \bibfield  {author} {\bibinfo {author} {\bibfnamefont {A.}~\bibnamefont
  {Giri}}, \bibinfo {author} {\bibfnamefont {N.}~\bibnamefont {Biswas}},
  \bibinfo {author} {\bibfnamefont {D.~L.}\ \bibnamefont {Chase}}, \bibinfo
  {author} {\bibfnamefont {N.}~\bibnamefont {Xue}}, \bibinfo {author}
  {\bibfnamefont {M.}~\bibnamefont {Abkarian}}, \bibinfo {author}
  {\bibfnamefont {S.}~\bibnamefont {Mendez}}, \bibinfo {author} {\bibfnamefont
  {S.}~\bibnamefont {Saha}},\ and\ \bibinfo {author} {\bibfnamefont {H.~A.}\
  \bibnamefont {Stone}},\ }\bibfield  {title} {\bibinfo {title} {Colliding
  respiratory jets as a mechanism of air exchange and pathogen transport during
  conversations},\ }\href {https://doi.org/10.1017/jfm.2021.915} {\bibfield
  {journal} {\bibinfo  {journal} {Journal of Fluid Mechanics}\ }\textbf
  {\bibinfo {volume} {930}},\ \bibinfo {pages} {R1} (\bibinfo {year}
  {2022})}\BibitemShut {NoStop}%
\bibitem [{\citenamefont {Bourouiba}(2021{\natexlab{a}})}]{bourouiba2021}%
  \BibitemOpen
  \bibfield  {author} {\bibinfo {author} {\bibfnamefont {L.}~\bibnamefont
  {Bourouiba}},\ }\bibfield  {title} {\bibinfo {title} {The fluid dynamics of
  disease transmission},\ }\href
  {https://doi.org/10.1146/annurev-fluid-060220-113712} {\bibfield  {journal}
  {\bibinfo  {journal} {Annual Review of Fluid Mechanics}\ }\textbf {\bibinfo
  {volume} {53}},\ \bibinfo {pages} {473} (\bibinfo {year}
  {2021}{\natexlab{a}})}\BibitemShut {NoStop}%
\bibitem [{\citenamefont {Bourouiba}(2021{\natexlab{b}})}]{bourouiba2021b}%
  \BibitemOpen
  \bibfield  {author} {\bibinfo {author} {\bibfnamefont {L.}~\bibnamefont
  {Bourouiba}},\ }\bibfield  {title} {\bibinfo {title} {Fluid dynamics of
  respiratory infectious diseases},\ }\href
  {https://doi.org/10.1146/annurev-bioeng-111820-025044} {\bibfield  {journal}
  {\bibinfo  {journal} {Annual Review of Biomedical Engineering}\ }\textbf
  {\bibinfo {volume} {23}},\ \bibinfo {pages} {547} (\bibinfo {year}
  {2021}{\natexlab{b}})},\ \bibinfo {note} {pMID: 34255991}\BibitemShut
  {NoStop}%
\bibitem [{\citenamefont {Morawska}\ \emph {et~al.}(2024)\citenamefont
  {Morawska}, \citenamefont {Allen}, \citenamefont {Bahnfleth}, \citenamefont
  {Bennett}, \citenamefont {Bluyssen}, \citenamefont {Boerstra}, \citenamefont
  {Buonanno}, \citenamefont {Cao}, \citenamefont {Dancer}, \citenamefont
  {Floto}, \citenamefont {Franchimon}, \citenamefont {Greenhalgh},
  \citenamefont {Haworth}, \citenamefont {Hogeling}, \citenamefont {Isaxon},
  \citenamefont {Jimenez}, \citenamefont {Kennedy}, \citenamefont {Kumar},
  \citenamefont {Kurnitski}, \citenamefont {Li}, \citenamefont {Loomans},
  \citenamefont {Marks}, \citenamefont {Marr}, \citenamefont {Mazzarella},
  \citenamefont {Melikov}, \citenamefont {Miller}, \citenamefont {Milton},
  \citenamefont {Monty}, \citenamefont {Nielsen}, \citenamefont {Noakes},
  \citenamefont {Peccia}, \citenamefont {Prather}, \citenamefont {Querol},
  \citenamefont {Salthammer}, \citenamefont {Sekhar}, \citenamefont
  {Seppänen}, \citenamefont {ichi Tanabe}, \citenamefont {Tang}, \citenamefont
  {Tellier}, \citenamefont {Tham}, \citenamefont {Wargocki}, \citenamefont
  {Wierzbicka},\ and\ \citenamefont {Yao}}]{morawska2024}%
  \BibitemOpen
  \bibfield  {author} {\bibinfo {author} {\bibfnamefont {L.}~\bibnamefont
  {Morawska}}, \bibinfo {author} {\bibfnamefont {J.}~\bibnamefont {Allen}},
  \bibinfo {author} {\bibfnamefont {W.}~\bibnamefont {Bahnfleth}}, \bibinfo
  {author} {\bibfnamefont {B.}~\bibnamefont {Bennett}}, \bibinfo {author}
  {\bibfnamefont {P.~M.}\ \bibnamefont {Bluyssen}}, \bibinfo {author}
  {\bibfnamefont {A.}~\bibnamefont {Boerstra}}, \bibinfo {author}
  {\bibfnamefont {G.}~\bibnamefont {Buonanno}}, \bibinfo {author}
  {\bibfnamefont {J.}~\bibnamefont {Cao}}, \bibinfo {author} {\bibfnamefont
  {S.~J.}\ \bibnamefont {Dancer}}, \bibinfo {author} {\bibfnamefont
  {A.}~\bibnamefont {Floto}}, \bibinfo {author} {\bibfnamefont
  {F.}~\bibnamefont {Franchimon}}, \bibinfo {author} {\bibfnamefont
  {T.}~\bibnamefont {Greenhalgh}}, \bibinfo {author} {\bibfnamefont
  {C.}~\bibnamefont {Haworth}}, \bibinfo {author} {\bibfnamefont
  {J.}~\bibnamefont {Hogeling}}, \bibinfo {author} {\bibfnamefont
  {C.}~\bibnamefont {Isaxon}}, \bibinfo {author} {\bibfnamefont {J.~L.}\
  \bibnamefont {Jimenez}}, \bibinfo {author} {\bibfnamefont {A.}~\bibnamefont
  {Kennedy}}, \bibinfo {author} {\bibfnamefont {P.}~\bibnamefont {Kumar}},
  \bibinfo {author} {\bibfnamefont {J.}~\bibnamefont {Kurnitski}}, \bibinfo
  {author} {\bibfnamefont {Y.}~\bibnamefont {Li}}, \bibinfo {author}
  {\bibfnamefont {M.}~\bibnamefont {Loomans}}, \bibinfo {author} {\bibfnamefont
  {G.}~\bibnamefont {Marks}}, \bibinfo {author} {\bibfnamefont {L.~C.}\
  \bibnamefont {Marr}}, \bibinfo {author} {\bibfnamefont {L.}~\bibnamefont
  {Mazzarella}}, \bibinfo {author} {\bibfnamefont {A.~K.}\ \bibnamefont
  {Melikov}}, \bibinfo {author} {\bibfnamefont {S.~L.}\ \bibnamefont {Miller}},
  \bibinfo {author} {\bibfnamefont {D.~K.}\ \bibnamefont {Milton}}, \bibinfo
  {author} {\bibfnamefont {J.}~\bibnamefont {Monty}}, \bibinfo {author}
  {\bibfnamefont {P.~V.}\ \bibnamefont {Nielsen}}, \bibinfo {author}
  {\bibfnamefont {C.}~\bibnamefont {Noakes}}, \bibinfo {author} {\bibfnamefont
  {J.}~\bibnamefont {Peccia}}, \bibinfo {author} {\bibfnamefont {K.~A.}\
  \bibnamefont {Prather}}, \bibinfo {author} {\bibfnamefont {X.}~\bibnamefont
  {Querol}}, \bibinfo {author} {\bibfnamefont {T.}~\bibnamefont {Salthammer}},
  \bibinfo {author} {\bibfnamefont {C.}~\bibnamefont {Sekhar}}, \bibinfo
  {author} {\bibfnamefont {O.}~\bibnamefont {Seppänen}}, \bibinfo {author}
  {\bibfnamefont {S.}~\bibnamefont {ichi Tanabe}}, \bibinfo {author}
  {\bibfnamefont {J.~W.}\ \bibnamefont {Tang}}, \bibinfo {author}
  {\bibfnamefont {R.}~\bibnamefont {Tellier}}, \bibinfo {author} {\bibfnamefont
  {K.~W.}\ \bibnamefont {Tham}}, \bibinfo {author} {\bibfnamefont
  {P.}~\bibnamefont {Wargocki}}, \bibinfo {author} {\bibfnamefont
  {A.}~\bibnamefont {Wierzbicka}},\ and\ \bibinfo {author} {\bibfnamefont
  {M.}~\bibnamefont {Yao}},\ }\bibfield  {title} {\bibinfo {title} {Mandating
  indoor air quality for public buildings},\ }\href
  {https://doi.org/10.1126/science.adl0677} {\bibfield  {journal} {\bibinfo
  {journal} {Science}\ }\textbf {\bibinfo {volume} {383}},\ \bibinfo {pages}
  {1418} (\bibinfo {year} {2024})}\BibitemShut {NoStop}%
\bibitem [{\citenamefont {Oswin}\ \emph {et~al.}(2022)\citenamefont {Oswin},
  \citenamefont {Haddrell}, \citenamefont {Otero-Fernandez}, \citenamefont
  {Mann}, \citenamefont {Cogan}, \citenamefont {Hilditch}, \citenamefont
  {Tian}, \citenamefont {Hardy}, \citenamefont {Hill}, \citenamefont {Finn}
  \emph {et~al.}}]{oswin2022}%
  \BibitemOpen
  \bibfield  {author} {\bibinfo {author} {\bibfnamefont {H.~P.}\ \bibnamefont
  {Oswin}}, \bibinfo {author} {\bibfnamefont {A.~E.}\ \bibnamefont {Haddrell}},
  \bibinfo {author} {\bibfnamefont {M.}~\bibnamefont {Otero-Fernandez}},
  \bibinfo {author} {\bibfnamefont {J.~F.}\ \bibnamefont {Mann}}, \bibinfo
  {author} {\bibfnamefont {T.~A.}\ \bibnamefont {Cogan}}, \bibinfo {author}
  {\bibfnamefont {T.~G.}\ \bibnamefont {Hilditch}}, \bibinfo {author}
  {\bibfnamefont {J.}~\bibnamefont {Tian}}, \bibinfo {author} {\bibfnamefont
  {D.~A.}\ \bibnamefont {Hardy}}, \bibinfo {author} {\bibfnamefont {D.~J.}\
  \bibnamefont {Hill}}, \bibinfo {author} {\bibfnamefont {A.}~\bibnamefont
  {Finn}}, \emph {et~al.},\ }\bibfield  {title} {\bibinfo {title} {{The
  dynamics of SARS-CoV-2 infectivity with changes in aerosol
  microenvironment}},\ }\href
  {https://doi.org/https://doi.org/10.1073/pnas.2200109119} {\bibfield
  {journal} {\bibinfo  {journal} {Proceedings National Academy of Sciences}\
  }\textbf {\bibinfo {volume} {119}},\ \bibinfo {pages} {e2200109119} (\bibinfo
  {year} {2022})}\BibitemShut {NoStop}%
\bibitem [{\citenamefont {Haddrell}\ \emph {et~al.}(2023)\citenamefont
  {Haddrell}, \citenamefont {Otero-Fernandez}, \citenamefont {Oswin},
  \citenamefont {Cogan}, \citenamefont {Bazire}, \citenamefont {Tian},
  \citenamefont {Alexander}, \citenamefont {Mann}, \citenamefont {Hill},
  \citenamefont {Finn} \emph {et~al.}}]{haddrell2023}%
  \BibitemOpen
  \bibfield  {author} {\bibinfo {author} {\bibfnamefont {A.}~\bibnamefont
  {Haddrell}}, \bibinfo {author} {\bibfnamefont {M.}~\bibnamefont
  {Otero-Fernandez}}, \bibinfo {author} {\bibfnamefont {H.}~\bibnamefont
  {Oswin}}, \bibinfo {author} {\bibfnamefont {T.}~\bibnamefont {Cogan}},
  \bibinfo {author} {\bibfnamefont {J.}~\bibnamefont {Bazire}}, \bibinfo
  {author} {\bibfnamefont {J.}~\bibnamefont {Tian}}, \bibinfo {author}
  {\bibfnamefont {R.}~\bibnamefont {Alexander}}, \bibinfo {author}
  {\bibfnamefont {J.~F.}\ \bibnamefont {Mann}}, \bibinfo {author}
  {\bibfnamefont {D.}~\bibnamefont {Hill}}, \bibinfo {author} {\bibfnamefont
  {A.}~\bibnamefont {Finn}}, \emph {et~al.},\ }\bibfield  {title} {\bibinfo
  {title} {{Differences in airborne stability of SARS-CoV-2 variants of concern
  is impacted by alkalinity of surrogates of respiratory aerosol}},\ }\href
  {https://doi.org/https://doi.org/10.1098/rsif.2023.0062} {\bibfield
  {journal} {\bibinfo  {journal} {J Royal Society Interface}\ }\textbf
  {\bibinfo {volume} {20}},\ \bibinfo {pages} {20230062} (\bibinfo {year}
  {2023})}\BibitemShut {NoStop}%
\bibitem [{\citenamefont {Augusto}\ \emph {et~al.}(2023)\citenamefont
  {Augusto}, \citenamefont {Murdolo}, \citenamefont {Chatzileontiadou},
  \citenamefont {Sabatino~Jr}, \citenamefont {Yusufali}, \citenamefont
  {Peyser}, \citenamefont {Butcher}, \citenamefont {Kizer}, \citenamefont
  {Guthrie}, \citenamefont {Murray} \emph {et~al.}}]{augusto2023}%
  \BibitemOpen
  \bibfield  {author} {\bibinfo {author} {\bibfnamefont {D.~G.}\ \bibnamefont
  {Augusto}}, \bibinfo {author} {\bibfnamefont {L.~D.}\ \bibnamefont
  {Murdolo}}, \bibinfo {author} {\bibfnamefont {D.~S.}\ \bibnamefont
  {Chatzileontiadou}}, \bibinfo {author} {\bibfnamefont {J.~J.}\ \bibnamefont
  {Sabatino~Jr}}, \bibinfo {author} {\bibfnamefont {T.}~\bibnamefont
  {Yusufali}}, \bibinfo {author} {\bibfnamefont {N.~D.}\ \bibnamefont
  {Peyser}}, \bibinfo {author} {\bibfnamefont {X.}~\bibnamefont {Butcher}},
  \bibinfo {author} {\bibfnamefont {K.}~\bibnamefont {Kizer}}, \bibinfo
  {author} {\bibfnamefont {K.}~\bibnamefont {Guthrie}}, \bibinfo {author}
  {\bibfnamefont {V.~W.}\ \bibnamefont {Murray}}, \emph {et~al.},\ }\bibfield
  {title} {\bibinfo {title} {{A common allele of HLA is associated with
  asymptomatic SARS-CoV-2 infection}},\ }\href
  {https://doi.org/https://doi.org/10.1038/s41586-023-06331-x} {\bibfield
  {journal} {\bibinfo  {journal} {Nature}\ }\textbf {\bibinfo {volume} {620}},\
  \bibinfo {pages} {128} (\bibinfo {year} {2023})}\BibitemShut {NoStop}%
\bibitem [{\citenamefont {Hill}(2012)}]{hill2012}%
  \BibitemOpen
  \bibfield  {author} {\bibinfo {author} {\bibfnamefont {A.~V.}\ \bibnamefont
  {Hill}},\ }\bibfield  {title} {\bibinfo {title} {Evolution, revolution and
  heresy in the genetics of infectious disease susceptibility},\ }\href
  {https://doi.org/https://doi.org/10.1098/rstb.2011.0275} {\bibfield
  {journal} {\bibinfo  {journal} {Philosophical Transactions of the Royal
  Society B: Biological Sciences}\ }\textbf {\bibinfo {volume} {367}},\
  \bibinfo {pages} {840} (\bibinfo {year} {2012})}\BibitemShut {NoStop}%
\bibitem [{\citenamefont {van~der Made}\ \emph {et~al.}(2022)\citenamefont
  {van~der Made}, \citenamefont {Netea}, \citenamefont {van~der Veerdonk},\
  and\ \citenamefont {Hoischen}}]{vandermade2022}%
  \BibitemOpen
  \bibfield  {author} {\bibinfo {author} {\bibfnamefont {C.~I.}\ \bibnamefont
  {van~der Made}}, \bibinfo {author} {\bibfnamefont {M.~G.}\ \bibnamefont
  {Netea}}, \bibinfo {author} {\bibfnamefont {F.~L.}\ \bibnamefont {van~der
  Veerdonk}},\ and\ \bibinfo {author} {\bibfnamefont {A.}~\bibnamefont
  {Hoischen}},\ }\bibfield  {title} {\bibinfo {title} {Clinical implications of
  host genetic variation and susceptibility to severe or critical covid-19},\
  }\href {https://doi.org/https://doi.org/10.1186/s13073-022-01100-3}
  {\bibfield  {journal} {\bibinfo  {journal} {Genome Medicine}\ }\textbf
  {\bibinfo {volume} {14}},\ \bibinfo {pages} {1} (\bibinfo {year}
  {2022})}\BibitemShut {NoStop}%
\bibitem [{\citenamefont {Viner}\ \emph {et~al.}(2021)\citenamefont {Viner},
  \citenamefont {Mytton}, \citenamefont {Bonell}, \citenamefont
  {Melendez-Torres}, \citenamefont {Ward}, \citenamefont {Hudson},
  \citenamefont {Waddington}, \citenamefont {Thomas}, \citenamefont {Russell},
  \citenamefont {van~der Klis}, \citenamefont {Koirala}, \citenamefont
  {Ladhani}, \citenamefont {Panovska-Griffiths}, \citenamefont {Davies},
  \citenamefont {Booy},\ and\ \citenamefont {Eggo}}]{viner2021}%
  \BibitemOpen
  \bibfield  {author} {\bibinfo {author} {\bibfnamefont {R.~M.}\ \bibnamefont
  {Viner}}, \bibinfo {author} {\bibfnamefont {O.~T.}\ \bibnamefont {Mytton}},
  \bibinfo {author} {\bibfnamefont {C.}~\bibnamefont {Bonell}}, \bibinfo
  {author} {\bibfnamefont {G.~J.}\ \bibnamefont {Melendez-Torres}}, \bibinfo
  {author} {\bibfnamefont {J.}~\bibnamefont {Ward}}, \bibinfo {author}
  {\bibfnamefont {L.}~\bibnamefont {Hudson}}, \bibinfo {author} {\bibfnamefont
  {C.}~\bibnamefont {Waddington}}, \bibinfo {author} {\bibfnamefont
  {J.}~\bibnamefont {Thomas}}, \bibinfo {author} {\bibfnamefont
  {S.}~\bibnamefont {Russell}}, \bibinfo {author} {\bibfnamefont
  {F.}~\bibnamefont {van~der Klis}}, \bibinfo {author} {\bibfnamefont
  {A.}~\bibnamefont {Koirala}}, \bibinfo {author} {\bibfnamefont
  {S.}~\bibnamefont {Ladhani}}, \bibinfo {author} {\bibfnamefont
  {J.}~\bibnamefont {Panovska-Griffiths}}, \bibinfo {author} {\bibfnamefont
  {N.~G.}\ \bibnamefont {Davies}}, \bibinfo {author} {\bibfnamefont
  {R.}~\bibnamefont {Booy}},\ and\ \bibinfo {author} {\bibfnamefont {R.~M.}\
  \bibnamefont {Eggo}},\ }\bibfield  {title} {\bibinfo {title} {{Susceptibility
  to SARS-CoV-2 Infection Among Children and Adolescents Compared With Adults:
  A Systematic Review and Meta-analysis}},\ }\href
  {https://doi.org/10.1001/jamapediatrics.2020.4573} {\bibfield  {journal}
  {\bibinfo  {journal} {JAMA Pediatrics}\ }\textbf {\bibinfo {volume} {175}},\
  \bibinfo {pages} {143} (\bibinfo {year} {2021})}\BibitemShut {NoStop}%
\bibitem [{\citenamefont {Shoham}\ \emph {et~al.}(2023)\citenamefont {Shoham},
  \citenamefont {Batista}, \citenamefont {Amor}, \citenamefont {Ergonul},
  \citenamefont {Hassanain}, \citenamefont {Hotez}, \citenamefont {Kang},
  \citenamefont {Kim}, \citenamefont {Lall}, \citenamefont {Larson} \emph
  {et~al.}}]{shoham2023}%
  \BibitemOpen
  \bibfield  {author} {\bibinfo {author} {\bibfnamefont {S.}~\bibnamefont
  {Shoham}}, \bibinfo {author} {\bibfnamefont {C.}~\bibnamefont {Batista}},
  \bibinfo {author} {\bibfnamefont {Y.~B.}\ \bibnamefont {Amor}}, \bibinfo
  {author} {\bibfnamefont {O.}~\bibnamefont {Ergonul}}, \bibinfo {author}
  {\bibfnamefont {M.}~\bibnamefont {Hassanain}}, \bibinfo {author}
  {\bibfnamefont {P.}~\bibnamefont {Hotez}}, \bibinfo {author} {\bibfnamefont
  {G.}~\bibnamefont {Kang}}, \bibinfo {author} {\bibfnamefont {J.~H.}\
  \bibnamefont {Kim}}, \bibinfo {author} {\bibfnamefont {B.}~\bibnamefont
  {Lall}}, \bibinfo {author} {\bibfnamefont {H.~J.}\ \bibnamefont {Larson}},
  \emph {et~al.},\ }\bibfield  {title} {\bibinfo {title} {Vaccines and
  therapeutics for immunocompromised patients with covid-19},\ }\href
  {https://doi.org/https://doi.org/10.1016/j.eclinm.2023.101965} {\bibfield
  {journal} {\bibinfo  {journal} {{eClinicalMedicine}}\ }\textbf {\bibinfo
  {volume} {59}},\ \bibinfo {pages} {101965} (\bibinfo {year}
  {2023})}\BibitemShut {NoStop}%
\bibitem [{\citenamefont {Paris}(2023)}]{paris2023}%
  \BibitemOpen
  \bibfield  {author} {\bibinfo {author} {\bibfnamefont {R.}~\bibnamefont
  {Paris}},\ }\bibfield  {title} {\bibinfo {title} {{SARS-CoV-2 Infection and
  Response to COVID-19 Vaccination in Patients With Primary
  Immunodeficiencies}},\ }\href {https://doi.org/10.1093/infdis/jiad145}
  {\bibfield  {journal} {\bibinfo  {journal} {Journal Infectious Diseases}\
  }\textbf {\bibinfo {volume} {228}},\ \bibinfo {pages} {S24} (\bibinfo {year}
  {2023})}\BibitemShut {NoStop}%
\bibitem [{\citenamefont {Kaplonek}\ \emph {et~al.}(2023)\citenamefont
  {Kaplonek}, \citenamefont {Cizmeci}, \citenamefont {Kwatra}, \citenamefont
  {Izu}, \citenamefont {Lee}, \citenamefont {Bertera}, \citenamefont
  {Fischinger}, \citenamefont {Mann}, \citenamefont {Amanat}, \citenamefont
  {Wang} \emph {et~al.}}]{kaplonek2023}%
  \BibitemOpen
  \bibfield  {author} {\bibinfo {author} {\bibfnamefont {P.}~\bibnamefont
  {Kaplonek}}, \bibinfo {author} {\bibfnamefont {D.}~\bibnamefont {Cizmeci}},
  \bibinfo {author} {\bibfnamefont {G.}~\bibnamefont {Kwatra}}, \bibinfo
  {author} {\bibfnamefont {A.}~\bibnamefont {Izu}}, \bibinfo {author}
  {\bibfnamefont {J.~S.-L.}\ \bibnamefont {Lee}}, \bibinfo {author}
  {\bibfnamefont {H.~L.}\ \bibnamefont {Bertera}}, \bibinfo {author}
  {\bibfnamefont {S.}~\bibnamefont {Fischinger}}, \bibinfo {author}
  {\bibfnamefont {C.}~\bibnamefont {Mann}}, \bibinfo {author} {\bibfnamefont
  {F.}~\bibnamefont {Amanat}}, \bibinfo {author} {\bibfnamefont
  {W.}~\bibnamefont {Wang}}, \emph {et~al.},\ }\bibfield  {title} {\bibinfo
  {title} {Chadox1 ncov-19 (azd1222) vaccine-induced fc receptor binding tracks
  with differential susceptibility to covid-19},\ }\href
  {https://doi.org/https://doi.org/10.1038/s41590-023-01513-1} {\bibfield
  {journal} {\bibinfo  {journal} {Nature Immunology}\ }\textbf {\bibinfo
  {volume} {24}},\ \bibinfo {pages} {1161} (\bibinfo {year}
  {2023})}\BibitemShut {NoStop}%
\bibitem [{\citenamefont {Chen}\ \emph {et~al.}(2021)\citenamefont {Chen},
  \citenamefont {Koopmans}, \citenamefont {Fisman},\ and\ \citenamefont
  {Gu}}]{chen2021}%
  \BibitemOpen
  \bibfield  {author} {\bibinfo {author} {\bibfnamefont {P.~Z.}\ \bibnamefont
  {Chen}}, \bibinfo {author} {\bibfnamefont {M.}~\bibnamefont {Koopmans}},
  \bibinfo {author} {\bibfnamefont {D.~N.}\ \bibnamefont {Fisman}},\ and\
  \bibinfo {author} {\bibfnamefont {F.~X.}\ \bibnamefont {Gu}},\ }\bibfield
  {title} {\bibinfo {title} {{Understanding why superspreading drives the
  COVID-19 pandemic but not the H1N1 pandemic}},\ }\href
  {https://doi.org/https://doi.org/10.1016/S1473-3099(21)00406-0} {\bibfield
  {journal} {\bibinfo  {journal} {The Lancet Infectious Diseases}\ }\textbf
  {\bibinfo {volume} {21}},\ \bibinfo {pages} {1203} (\bibinfo {year}
  {2021})}\BibitemShut {NoStop}%
\bibitem [{\citenamefont {Wegehaupt}\ \emph {et~al.}(2023)\citenamefont
  {Wegehaupt}, \citenamefont {Endo},\ and\ \citenamefont
  {Vassall}}]{wegehaupt2023}%
  \BibitemOpen
  \bibfield  {author} {\bibinfo {author} {\bibfnamefont {O.}~\bibnamefont
  {Wegehaupt}}, \bibinfo {author} {\bibfnamefont {A.}~\bibnamefont {Endo}},\
  and\ \bibinfo {author} {\bibfnamefont {A.}~\bibnamefont {Vassall}},\
  }\bibfield  {title} {\bibinfo {title} {{Superspreading, overdispersion and
  their implications in the SARS-CoV-2 (COVID-19) pandemic: a systematic review
  and meta-analysis of the literature}},\ }\href
  {https://doi.org/https://doi.org/10.1186/s12889-023-15915-1} {\bibfield
  {journal} {\bibinfo  {journal} {BMC Public Health}\ }\textbf {\bibinfo
  {volume} {23}},\ \bibinfo {pages} {1} (\bibinfo {year} {2023})}\BibitemShut
  {NoStop}%
\bibitem [{\citenamefont {Mossong}\ \emph {et~al.}(2008)\citenamefont
  {Mossong}, \citenamefont {Hens}, \citenamefont {Jit}, \citenamefont
  {Beutels}, \citenamefont {Auranen}, \citenamefont {Mikolajczyk},
  \citenamefont {Massari}, \citenamefont {Salmaso}, \citenamefont {Tomba},
  \citenamefont {Wallinga}, \citenamefont {Heijne}, \citenamefont
  {Sadkowska-Todys}, \citenamefont {Rosinska},\ and\ \citenamefont
  {Edmunds}}]{mossong2008}%
  \BibitemOpen
  \bibfield  {author} {\bibinfo {author} {\bibfnamefont {J.}~\bibnamefont
  {Mossong}}, \bibinfo {author} {\bibfnamefont {N.}~\bibnamefont {Hens}},
  \bibinfo {author} {\bibfnamefont {M.}~\bibnamefont {Jit}}, \bibinfo {author}
  {\bibfnamefont {P.}~\bibnamefont {Beutels}}, \bibinfo {author} {\bibfnamefont
  {K.}~\bibnamefont {Auranen}}, \bibinfo {author} {\bibfnamefont
  {R.}~\bibnamefont {Mikolajczyk}}, \bibinfo {author} {\bibfnamefont
  {M.}~\bibnamefont {Massari}}, \bibinfo {author} {\bibfnamefont
  {S.}~\bibnamefont {Salmaso}}, \bibinfo {author} {\bibfnamefont {G.~S.}\
  \bibnamefont {Tomba}}, \bibinfo {author} {\bibfnamefont {J.}~\bibnamefont
  {Wallinga}}, \bibinfo {author} {\bibfnamefont {J.}~\bibnamefont {Heijne}},
  \bibinfo {author} {\bibfnamefont {M.}~\bibnamefont {Sadkowska-Todys}},
  \bibinfo {author} {\bibfnamefont {M.}~\bibnamefont {Rosinska}},\ and\
  \bibinfo {author} {\bibfnamefont {W.~J.}\ \bibnamefont {Edmunds}},\
  }\bibfield  {title} {\bibinfo {title} {Social contacts and mixing patterns
  relevant to the spread of infectious diseases},\ }\href
  {https://doi.org/10.1371/journal.pmed.0050074} {\bibfield  {journal}
  {\bibinfo  {journal} {PLOS Medicine}\ }\textbf {\bibinfo {volume} {5}},\
  \bibinfo {pages} {1} (\bibinfo {year} {2008})}\BibitemShut {NoStop}%
\bibitem [{\citenamefont {Stumpf}\ and\ \citenamefont
  {Porter}(2012)}]{stumpf2012}%
  \BibitemOpen
  \bibfield  {author} {\bibinfo {author} {\bibfnamefont {M.~P.~H.}\
  \bibnamefont {Stumpf}}\ and\ \bibinfo {author} {\bibfnamefont {M.~A.}\
  \bibnamefont {Porter}},\ }\bibfield  {title} {\bibinfo {title} {Critical
  truths about power laws},\ }\href {https://doi.org/10.1126/science.1216142}
  {\bibfield  {journal} {\bibinfo  {journal} {Science}\ }\textbf {\bibinfo
  {volume} {335}},\ \bibinfo {pages} {665} (\bibinfo {year}
  {2012})}\BibitemShut {NoStop}%
\bibitem [{\citenamefont {Kissler}\ \emph {et~al.}(2021)\citenamefont
  {Kissler}, \citenamefont {Fauver}, \citenamefont {Mack}, \citenamefont
  {Olesen}, \citenamefont {Tai}, \citenamefont {Shiue}, \citenamefont
  {Kalinich}, \citenamefont {Jednak}, \citenamefont {Ott}, \citenamefont
  {Vogels} \emph {et~al.}}]{kissler2021}%
  \BibitemOpen
  \bibfield  {author} {\bibinfo {author} {\bibfnamefont {S.~M.}\ \bibnamefont
  {Kissler}}, \bibinfo {author} {\bibfnamefont {J.~R.}\ \bibnamefont {Fauver}},
  \bibinfo {author} {\bibfnamefont {C.}~\bibnamefont {Mack}}, \bibinfo {author}
  {\bibfnamefont {S.~W.}\ \bibnamefont {Olesen}}, \bibinfo {author}
  {\bibfnamefont {C.}~\bibnamefont {Tai}}, \bibinfo {author} {\bibfnamefont
  {K.~Y.}\ \bibnamefont {Shiue}}, \bibinfo {author} {\bibfnamefont {C.~C.}\
  \bibnamefont {Kalinich}}, \bibinfo {author} {\bibfnamefont {S.}~\bibnamefont
  {Jednak}}, \bibinfo {author} {\bibfnamefont {I.~M.}\ \bibnamefont {Ott}},
  \bibinfo {author} {\bibfnamefont {C.~B.}\ \bibnamefont {Vogels}}, \emph
  {et~al.},\ }\bibfield  {title} {\bibinfo {title} {{Viral dynamics of acute
  SARS-CoV-2 infection and applications to diagnostic and public health
  strategies}},\ }\href
  {https://doi.org/https://doi.org/10.1371/journal.pbio.3001333} {\bibfield
  {journal} {\bibinfo  {journal} {PLoS biology}\ }\textbf {\bibinfo {volume}
  {19}},\ \bibinfo {pages} {e3001333} (\bibinfo {year} {2021})}\BibitemShut
  {NoStop}%
\bibitem [{\citenamefont {Proschan}(1963)}]{proschan1963}%
  \BibitemOpen
  \bibfield  {author} {\bibinfo {author} {\bibfnamefont {F.}~\bibnamefont
  {Proschan}},\ }\bibfield  {title} {\bibinfo {title} {Theoretical explanation
  of observed decreasing failure rate},\ }\href
  {https://doi.org/10.1080/00401706.1963.10490105} {\bibfield  {journal}
  {\bibinfo  {journal} {Technometrics}\ }\textbf {\bibinfo {volume} {5}},\
  \bibinfo {pages} {375} (\bibinfo {year} {1963})}\BibitemShut {NoStop}%
\bibitem [{\citenamefont {Oberg}\ and\ \citenamefont
  {Brosseau}(2008)}]{oberg2008}%
  \BibitemOpen
  \bibfield  {author} {\bibinfo {author} {\bibfnamefont {T.}~\bibnamefont
  {Oberg}}\ and\ \bibinfo {author} {\bibfnamefont {L.~M.}\ \bibnamefont
  {Brosseau}},\ }\bibfield  {title} {\bibinfo {title} {Surgical mask filter and
  fit performance},\ }\href {https://doi.org/10.1016/j.ajic.2007.07.008}
  {\bibfield  {journal} {\bibinfo  {journal} {American Journal of Infection
  Control}\ }\textbf {\bibinfo {volume} {36}},\ \bibinfo {pages} {276}
  (\bibinfo {year} {2008})}\BibitemShut {NoStop}%
\bibitem [{\citenamefont {Vegvari}\ \emph {et~al.}(2022)\citenamefont
  {Vegvari}, \citenamefont {Abbott}, \citenamefont {Ball}, \citenamefont
  {Brooks-Pollock}, \citenamefont {Challen}, \citenamefont {Collyer},
  \citenamefont {Dangerfield}, \citenamefont {Gog}, \citenamefont {Gostic},
  \citenamefont {Heffernan}, \citenamefont {Hollingsworth}, \citenamefont
  {Isham}, \citenamefont {Kenah}, \citenamefont {Mollison}, \citenamefont
  {Panovska-Griffiths}, \citenamefont {Pellis}, \citenamefont {Roberts},
  \citenamefont {Tomba}, \citenamefont {Thompson},\ and\ \citenamefont
  {Trapman}}]{vegvari2022}%
  \BibitemOpen
  \bibfield  {author} {\bibinfo {author} {\bibfnamefont {C.}~\bibnamefont
  {Vegvari}}, \bibinfo {author} {\bibfnamefont {S.}~\bibnamefont {Abbott}},
  \bibinfo {author} {\bibfnamefont {F.}~\bibnamefont {Ball}}, \bibinfo {author}
  {\bibfnamefont {E.}~\bibnamefont {Brooks-Pollock}}, \bibinfo {author}
  {\bibfnamefont {R.}~\bibnamefont {Challen}}, \bibinfo {author} {\bibfnamefont
  {B.~S.}\ \bibnamefont {Collyer}}, \bibinfo {author} {\bibfnamefont
  {C.}~\bibnamefont {Dangerfield}}, \bibinfo {author} {\bibfnamefont {J.~R.}\
  \bibnamefont {Gog}}, \bibinfo {author} {\bibfnamefont {K.~M.}\ \bibnamefont
  {Gostic}}, \bibinfo {author} {\bibfnamefont {J.~M.}\ \bibnamefont
  {Heffernan}}, \bibinfo {author} {\bibfnamefont {T.~D.}\ \bibnamefont
  {Hollingsworth}}, \bibinfo {author} {\bibfnamefont {V.}~\bibnamefont
  {Isham}}, \bibinfo {author} {\bibfnamefont {E.}~\bibnamefont {Kenah}},
  \bibinfo {author} {\bibfnamefont {D.}~\bibnamefont {Mollison}}, \bibinfo
  {author} {\bibfnamefont {J.}~\bibnamefont {Panovska-Griffiths}}, \bibinfo
  {author} {\bibfnamefont {L.}~\bibnamefont {Pellis}}, \bibinfo {author}
  {\bibfnamefont {M.~G.}\ \bibnamefont {Roberts}}, \bibinfo {author}
  {\bibfnamefont {G.~S.}\ \bibnamefont {Tomba}}, \bibinfo {author}
  {\bibfnamefont {R.~N.}\ \bibnamefont {Thompson}},\ and\ \bibinfo {author}
  {\bibfnamefont {P.}~\bibnamefont {Trapman}},\ }\bibfield  {title} {\bibinfo
  {title} {{Commentary on the use of the reproduction number R during the
  COVID-19 pandemic}},\ }\href {https://doi.org/10.1177/09622802211037079}
  {\bibfield  {journal} {\bibinfo  {journal} {Statistical Methods Medical
  Research}\ }\textbf {\bibinfo {volume} {31}},\ \bibinfo {pages} {1675}
  (\bibinfo {year} {2022})}\BibitemShut {NoStop}%
\bibitem [{\citenamefont {{van den Driessche}}(2017)}]{vanderdriessche2017}%
  \BibitemOpen
  \bibfield  {author} {\bibinfo {author} {\bibfnamefont {P.}~\bibnamefont {{van
  den Driessche}}},\ }\bibfield  {title} {\bibinfo {title} {Reproduction
  numbers of infectious disease models},\ }\href
  {https://doi.org/https://doi.org/10.1016/j.idm.2017.06.002} {\bibfield
  {journal} {\bibinfo  {journal} {Infectious Disease Modelling}\ }\textbf
  {\bibinfo {volume} {2}},\ \bibinfo {pages} {288} (\bibinfo {year}
  {2017})}\BibitemShut {NoStop}%
\bibitem [{\citenamefont {Cori}\ \emph {et~al.}(2013)\citenamefont {Cori},
  \citenamefont {Ferguson}, \citenamefont {Fraser},\ and\ \citenamefont
  {Cauchemez}}]{cori2013}%
  \BibitemOpen
  \bibfield  {author} {\bibinfo {author} {\bibfnamefont {A.}~\bibnamefont
  {Cori}}, \bibinfo {author} {\bibfnamefont {N.~M.}\ \bibnamefont {Ferguson}},
  \bibinfo {author} {\bibfnamefont {C.}~\bibnamefont {Fraser}},\ and\ \bibinfo
  {author} {\bibfnamefont {S.}~\bibnamefont {Cauchemez}},\ }\bibfield  {title}
  {\bibinfo {title} {{A New Framework and Software to Estimate Time-Varying
  Reproduction Numbers During Epidemics}},\ }\href
  {https://doi.org/10.1093/aje/kwt133} {\bibfield  {journal} {\bibinfo
  {journal} {American Journal Epidemiology}\ }\textbf {\bibinfo {volume}
  {178}},\ \bibinfo {pages} {1505} (\bibinfo {year} {2013})}\BibitemShut
  {NoStop}%
\bibitem [{\citenamefont {Boulos}\ \emph {et~al.}(2023)\citenamefont {Boulos},
  \citenamefont {Curran}, \citenamefont {Gallant}, \citenamefont {Wong},
  \citenamefont {Johnson}, \citenamefont {Delahunty-Pike}, \citenamefont
  {Saxinger}, \citenamefont {Chu}, \citenamefont {Comeau}, \citenamefont
  {Flynn}, \citenamefont {Clegg},\ and\ \citenamefont {Dye}}]{boulos2023}%
  \BibitemOpen
  \bibfield  {author} {\bibinfo {author} {\bibfnamefont {L.}~\bibnamefont
  {Boulos}}, \bibinfo {author} {\bibfnamefont {J.~A.}\ \bibnamefont {Curran}},
  \bibinfo {author} {\bibfnamefont {A.}~\bibnamefont {Gallant}}, \bibinfo
  {author} {\bibfnamefont {H.}~\bibnamefont {Wong}}, \bibinfo {author}
  {\bibfnamefont {C.}~\bibnamefont {Johnson}}, \bibinfo {author} {\bibfnamefont
  {A.}~\bibnamefont {Delahunty-Pike}}, \bibinfo {author} {\bibfnamefont
  {L.}~\bibnamefont {Saxinger}}, \bibinfo {author} {\bibfnamefont
  {D.}~\bibnamefont {Chu}}, \bibinfo {author} {\bibfnamefont {J.}~\bibnamefont
  {Comeau}}, \bibinfo {author} {\bibfnamefont {T.}~\bibnamefont {Flynn}},
  \bibinfo {author} {\bibfnamefont {J.}~\bibnamefont {Clegg}},\ and\ \bibinfo
  {author} {\bibfnamefont {C.}~\bibnamefont {Dye}},\ }\bibfield  {title}
  {\bibinfo {title} {{Effectiveness of face masks for reducing transmission of
  SARS-CoV-2: a rapid systematic review}},\ }\href
  {https://doi.org/10.1098/rsta.2023.0133} {\bibfield  {journal} {\bibinfo
  {journal} {Philosophical Transactions of the Royal Society A: Mathematical,
  Physical and Engineering Sciences}\ }\textbf {\bibinfo {volume} {381}},\
  \bibinfo {pages} {20230133} (\bibinfo {year} {2023})}\BibitemShut {NoStop}%
\bibitem [{\citenamefont {Bar-Yam}\ \emph {et~al.}(2023)\citenamefont
  {Bar-Yam}, \citenamefont {Samet}, \citenamefont {Siegenfeld},\ and\
  \citenamefont {Taleb}}]{baryam2023}%
  \BibitemOpen
  \bibfield  {author} {\bibinfo {author} {\bibfnamefont {Y.}~\bibnamefont
  {Bar-Yam}}, \bibinfo {author} {\bibfnamefont {J.~M.}\ \bibnamefont {Samet}},
  \bibinfo {author} {\bibfnamefont {A.~F.}\ \bibnamefont {Siegenfeld}},\ and\
  \bibinfo {author} {\bibfnamefont {N.~N.}\ \bibnamefont {Taleb}},\ }\href@noop
  {} {\bibinfo {title} {Quantitative errors in the cochrane review on "physical
  interventions to interrupt or reduce the spread of respiratory viruses"}}
  (\bibinfo {year} {2023}),\ \Eprint {https://arxiv.org/abs/2310.15198}
  {arXiv:2310.15198 [physics.soc-ph]} \BibitemShut {NoStop}%
\bibitem [{\citenamefont {Ke}\ \emph {et~al.}(2021)\citenamefont {Ke},
  \citenamefont {Zitzmann}, \citenamefont {Ho}, \citenamefont {Ribeiro},\ and\
  \citenamefont {Perelson}}]{ke2021}%
  \BibitemOpen
  \bibfield  {author} {\bibinfo {author} {\bibfnamefont {R.}~\bibnamefont
  {Ke}}, \bibinfo {author} {\bibfnamefont {C.}~\bibnamefont {Zitzmann}},
  \bibinfo {author} {\bibfnamefont {D.~D.}\ \bibnamefont {Ho}}, \bibinfo
  {author} {\bibfnamefont {R.~M.}\ \bibnamefont {Ribeiro}},\ and\ \bibinfo
  {author} {\bibfnamefont {A.~S.}\ \bibnamefont {Perelson}},\ }\bibfield
  {title} {\bibinfo {title} {In vivo kinetics of sars-cov-2 infection and its
  relationship with a person’s infectiousness},\ }\href
  {https://doi.org/https://doi.org/10.1073/pnas.2111477118} {\bibfield
  {journal} {\bibinfo  {journal} {Proceedings of the National Academy of
  Sciences}\ }\textbf {\bibinfo {volume} {118}},\ \bibinfo {pages}
  {e2111477118} (\bibinfo {year} {2021})}\BibitemShut {NoStop}%
\end{thebibliography}

%

\end{document}